\documentclass[showkeys,floatfix,aps,10pt,prd]{revtex4-2}
\usepackage{graphicx,epstopdf}
\usepackage[caption=false]{subfig}
\usepackage{appendix}
\usepackage[T1]{fontenc}
\usepackage{lmodern}
\usepackage[dvipsnames,x11names]{xcolor}
\usepackage[colorlinks=true,linkcolor=Magenta!80!black,citecolor=Magenta!70!black,urlcolor=Magenta!80!black]{hyperref}
\usepackage[sort&compress]{natbib}
\usepackage{morefloats}
\usepackage[pdf]{pstricks}
\usepackage{amsmath}
\usepackage{amssymb}
\usepackage{amsfonts}
\usepackage{rotating}
\usepackage{cancel}
\usepackage{mathtools}
\usepackage{bbm}
\usepackage{dsfont}
\usepackage{bbold}
\usepackage{multirow}
\usepackage{ulem}
\usepackage{physics}
\usepackage{orcidlink}
\usepackage{colortbl}
 \usepackage{booktabs}

\begin{document}

\title{Structural dissection of hadronic molecules: \\
       The $D^{(*)}\bar{K}^{(*)}$ family under QCD light-cone sum rules}

\author{Ula\c{s}~\"{O}zdem\orcidlink{0000-0002-1907-2894}}%
\email[]{ulasozdem@aydin.edu.tr }
\affiliation{ Health Services Vocational School of Higher Education, Istanbul Aydin University, Sefakoy-Kucukcekmece, 34295 Istanbul, 
T\"{u}rkiye}

 
\begin{abstract}
We investigate the static electromagnetic properties of three charm--strange molecular tetraquark candidates with quantum numbers $J^{P}=1^{+}$, namely the $D\bar{K}^{\ast}$, $D^{\ast}\bar{K}$, and $D^{\ast}\bar{K}^{\ast}$ systems. The analysis is carried out within the framework of QCD light-cone sum rules, using interpolating currents constructed from colour-singlet meson bilinears to reflect their molecular configurations. Both perturbative and non-perturbative photon contributions are included, and numerical predictions for the magnetic and electric quadrupole moments are obtained. The magnetic moments are found to lie in the range $1$--$3$ nuclear magnetons, with the largest value associated with the $D^{*}\bar{K}$ configuration. The quadrupole moments are an order of magnitude smaller, of order $10^{-3}\,\mathrm{fm}^{2}$, indicating only weak deviations from spherical charge distributions. A flavour decomposition shows that the magnetic response is dominated by the light quarks, while the charm-quark contribution is strongly suppressed, a feature naturally expected for loosely bound hadronic molecules. The present analysis extends QCD light-cone sum-rule studies of exotic hadrons by providing a systematic determination of the electromagnetic moments of the $D^{(\ast)}\bar K^{(\ast)}$ molecular systems. These results provide quantitative benchmarks and identify several qualitative fingerprints---most notably the vanishing electromagnetic moments of the neutral $D\bar K^{*}$, the strong charm-quark suppression, and the smallness of the quadrupole deformations---that, taken together, distinguish the molecular scenario from compact diquark--antidiquark interpretations and may offer useful guidance for future experimental studies of charm--strange exotic states.
\end{abstract}


\maketitle

\section{Introduction and Motivation}\label{motivation}

The discovery of exotic hadrons—states whose quark content defies the traditional meson ($q\bar{q}$) and baryon ($qqq$) classifications—has revolutionized our understanding of quantum chromodynamics (QCD). The journey began with the observation of the $X(3872)$ by Belle in 2003~\cite{Belle:2003nnu}, which was quickly confirmed by several experiments and could not be accommodated within the charmonium spectrum. Since then, a wealth of unconventional resonances, collectively labeled as $XYZ$ states, pentaquarks, and tetraquarks, has been reported at facilities worldwide, including LHCb, Belle, BESIII, CMS, and others. These discoveries have sparked intense theoretical activity, as standard quark‑model pictures are inadequate to describe their properties. Comprehensive reviews document the experimental progress and the diverse theoretical interpretations put forward to explain them~\cite{Esposito:2014rxa, Esposito:2016noz, Olsen:2017bmm, Lebed:2016hpi, Nielsen:2009uh, Brambilla:2019esw, Agaev:2020zad, Chen:2016qju, Ali:2017jda, Guo:2017jvc, Liu:2019zoy, Yang:2020atz, Dong:2021juy, Dong:2021bvy, Chen:2022asf, Meng:2022ozq, Wang:2025sic}.

A particularly active frontier is the charm‑strange sector, where LHCb has recently reported several tetraquark candidates. In analyses of $B^+ \to D^+ D^- K^+$ decays, two resonances—$T_{cs0}(2900)^0$ and $T_{cs1}(2900)^0$—were observed with minimal quark content $\bar{c}s u\bar{d}$~\cite{LHCb:2020pxc,LHCb:2020bls}. Their quantum numbers were determined to be $J^P=0^+$ and $J^P=1^-$, respectively. Subsequently, in studies of $B$ decays to $D_s\pi$ final states, LHCb identified two additional states, $T_{a\,cs}(2900)^0$ and $T_{a\,cs}(2900)^{++}$, forming an isospin triplet with $I(J^P)=1(0)^+$~\cite{LHCb:2022lzp,LHCb:2022sfr}.
In 2024, the LHCb Collaboration carried out an analysis of resonant structures in the decays $B^+ \to D^{\ast +} D^- K^+$ and $B^+ \to D^{\ast -} D^+ K^+$, using proton--proton collision data collected at centre-of-mass energies of $\sqrt{s}=7$, $8$, and 13~$\text{TeV}$~\cite{LHCb:2024vfz}. In addition to several charmonium-like states, including $\eta_c(3945)$, $h_c(4000)$, $\chi_{c1}(4010)$, and $h_c(4300)$,
the analysis confirmed the presence of the $T^{\ast}_{\bar{c}s0}(2870)^0$ and $T^{\ast}_{\bar{c}s1}(2900)^0$ resonances in the $D^-K^+$ invariant mass spectrum.
These structures, previously observed in the $B^+ \to D^+ D^- K^+$ decay channel, were thus established in an independent production mode, with measured masses and decay widths consistent with earlier observations. All these resonances lie near 2.9 GeV, firmly establishing the existence of four‑quark systems containing both charm and strange quarks.  While the experimentally established states carry $J^P=0^+$ or $1^-$, theoretical models predict that the same quark constituents should also form resonances with other quantum numbers, notably $J^P=1^+$ and $2^+$~\cite{Molina:2020hde, Dai:2022qwh, Sundu:2022kyd, Azizi:2021aib, Agaev:2021jsz, Mutuk:2020igv, Zhou:2025yjb, Zhou:2025rpb}. To elucidate the internal structure of these systems—whether they are compact tetraquarks, loosely bound meson‑molecule combinations, or other configurations—it is essential to study observables beyond masses and decay widths. Electromagnetic moments offer precisely such a probe: the magnetic dipole moment reflects the distribution of circulating currents, while the electric quadrupole moment  measures the deviation from a spherical charge distribution. These moments are sensitive to the spatial wave function and spin alignment of the constituents, providing distinctive fingerprints that can discriminate among different structural models.

In this work we compute the magnetic and quadrupole moments of three $J^P=1^+$ molecular candidates: $D\bar K^*$, $D^*\bar K$, and $D^*\bar K^*$. The calculation is performed within the framework of light‑cone sum rules (LCSR)~\cite{Chernyak:1990ag, Braun:1988qv, Balitsky:1989ry}, a method that combines QCD operator‑product expansion near the light cone with dispersion relations. The LCSR approach systematically incorporates both short‑distance (perturbative) photon couplings and long‑distance effects described by photon distribution amplitudes. It has been successfully applied to electromagnetic properties of  singly‑heavy exotic hadrons~\cite{Ozdem:2024ydl, Ozdem:2024pyb, Ozdem:2023edw, Ozdem:2023okg, Ozdem:2023eyz, Ozdem:2022ydv, Ozdem:2021vry, Azizi:2021aib, Azizi:2018jky, Azizi:2018mte, Ozdem:2025olj}, and here we extend it to charm‑strange tetraquark systems. Our results provide quantitative predictions for electromagnetic moments that can be confronted with future experimental data, for instance from photon‑induced production or from angular analyses of radiative decays. Moreover, the pattern of moments—especially the relative sizes of light‑quark and charm‑quark contributions—offers clues about the molecular or compact nature of the states. The present study thus complements earlier spectroscopic investigations and adds a new dimension to the characterization of exotic charm‑strange hadrons.

The paper is organized as follows. In Sec.~\ref{formalism} we introduce the LCSR formalism, construct the correlation functions appropriate for each molecular configuration, and derive the sum rules for the magnetic and quadrupole moments. Section~\ref{numerical} contains the numerical analysis: we specify the input parameters, determine the working windows of the sum‑rule auxiliary parameters, and present the resulting moments together with an error estimate. Finally, Sec.~\ref{sum} summarizes our findings and discusses their implications for future experimental searches and for the broader understanding of exotic hadron structure. Lengthy expressions for the spectral densities are relegated to the Appendix.

 \begin{widetext}
 
\section{Theoretical framework for electromagnetic properties}\label{formalism}

\subsection{Correlation function and external field formulation}

The electromagnetic properties of the charm-strange molecular tetraquark states (for short $T_{cs}$)—specifically their magnetic ($\mu$) and quadrupole ($\mathcal{D}$) moments—can be systematically extracted using LCSR. The analysis begins with the three-point correlation function
\begin{equation}
 \label{edmn00}
T_{\mu\nu\alpha}(p,q)=i^2\int d^{4}x\int d^{4}y\,e^{ip\cdot x+iq\cdot y}\,
\langle 0|\mathcal{T}\{J_{\mu}(x) J_{\alpha}^{em}(y)
J_{\nu}^{\dagger}(0)\}|0\rangle,    
\end{equation}
where $J_{\alpha}^{em}=\sum_{q}e_q\bar{q}\gamma_\alpha q$ is the electromagnetic current ($e_q$ being the quark electric charge), and $J_\mu(x)$ denotes an interpolating current that couples to the $J^P=1^+$ tetraquark state.  The correlator in Eq. (\ref{edmn00}) probes the matrix element $\langle T_{cs}(p)|J_{\alpha}^{em}|T_{cs}(p')\rangle$ through analytic continuation, where the initial and final tetraquark momenta are $p$ and $p'=p+q$, respectively.

In practice, it is technically advantageous to reformulate the problem using an external background electromagnetic field. Replacing the explicit photon insertion by a classical, slowly varying field $A_\mu^{\text{ext}}(x)$, one considers instead the two-point function
\begin{equation}
 \label{edmn01}
T_{\mu\nu}(p,q)=i\int d^{4}x\,e^{ip\cdot x}\,
\langle 0|\mathcal{T}\{J_{\mu}(x)J_{\nu}^{\dagger}(0)\}|0\rangle_{F}, 
\end{equation}
where the subscript $F$ indicates that the vacuum expectation value is evaluated in the presence of the external field. The field strength tensor is taken in momentum space as 
\begin{equation*}
F_{\alpha\beta}=i(\varepsilon_\alpha q_\beta-\varepsilon_\beta q_\alpha)e^{-iq\cdot x},
\end{equation*}
with $q$ and $\varepsilon$ being the four‑momentum and polarization vector of the background photon. 

Advantages of the external‑field method:  
(i) Gauge invariance is maintained manifestly; 
(ii) soft (non‑perturbative) and hard (perturbative) photon emissions are separated cleanly; 
(iii) the photon light‑cone distribution amplitudes (LCDAs) enter naturally, organizing the operator product expansion (OPE) in powers of the photon virtuality~\cite{Ball:2002ps}.

Because the external field is weak, the correlation function can be expanded as
\begin{equation}
T_{\mu\nu}(p,q) = T_{\mu\nu}^{(0)}(p) + T_{\mu\nu}^{(1)}(p,q) + \mathcal{O}(F^{2}),
\label{eq:expansion}
\end{equation}
where $T_{\mu\nu}^{(0)}$ corresponds to the free propagation of the tetraquark (giving rise to mass sum rules) and $T_{\mu\nu}^{(1)}$ is linear in the field strength. Only the latter term contains the information on electromagnetic multipole moments~\cite{Ball:2002ps,Novikov:1983gd,Ioffe:1983ju}.  The external-field formulation adopted in Eq.~(\ref{edmn01}) yields 
the static form factors at $Q^{2}=0$ but does not retain the explicit 
$Q^{2}$ dependence required to compute charge radii. Quantities such 
as $\langle r^{2} \rangle = 6\,\mathrm{d}F/\mathrm{d}Q^{2}\big|_{Q^{2}=0}$ 
follow from a full three-point analysis with an explicit photon 
insertion, as in Eq.~(\ref{edmn00}). Although technically more 
demanding, such a calculation would provide a direct numerical estimate of the spatial extension of these states and represents a natural extension of the present work.

The formulation outlined above allows us to compute the electromagnetic moments of the tetraquark states within the LCSR approach. The actual calculation proceeds in three consecutive steps:

\begin{itemize}
    \item Hadronic representation: Using dispersion relations, the correlation function is expressed in terms of the tetraquark's mass $m$, coupling $\lambda$ (defined by $\langle0|J_\mu|T_{cs}\rangle=\lambda\varepsilon_\mu$), and electromagnetic form factors. The contribution of the ground state is isolated, while higher resonances and the continuum are parametrized through a spectral density.
    
   \item QCD representation: The same correlation function is computed via the OPE near the light cone ($x^2\sim0$). The short‑distance part is treated perturbatively, while the long‑distance interaction of the photon with the QCD vacuum is encoded in photon LCDAs of increasing twist. This yields $T^{\text{QCD}}_{\mu\nu}$, written explicitly in terms of quark‑gluon degrees of freedom.
    
    \item Matching and Borel transformation: The two representations are matched assuming quark‑hadron duality. To suppress contributions from excited states and enhance the ground‑state pole, a double Borel transformation with respect to $-p^2$ and $- p^{\prime 2}$ is applied. After subtracting the continuum using a suitable threshold $s_0$, one obtains sum rules for the magnetic and quadrupole moments.
\end{itemize}

In the following sections, we first specify the interpolating currents for the $T_{cs}$ molecular states, then derive the hadronic and QCD sides in detail, and finally extract the sum rules for the electromagnetic moments.

\subsection{Interpolating currents for the $T_{cs}$ molecular states} \label{sec:IIB}

The construction of appropriate interpolating currents is a key step in the QCD sum-rule analysis of multiquark states, since the chosen operators determine which hadronic configurations are preferentially selected from the QCD vacuum. For the molecular-type $T_{cs}$ tetraquarks studied here, we employ currents that are explicitly built from colour-singlet meson bilinears. This reflects the underlying physical picture: the states are assumed to be dominantly composed of two weakly bound, colourless mesons. Such a construction emphasises the long-distance, molecular component of the wave function, in contrast to compact diquark–antidiquark currents, which would probe a more spatially concentrated configuration.
The interpolating currents for the three $T_{cs}$ molecular systems are defined as
\begin{align}
\label{curr1}
J_{\mu }^{1}(x) &=  [\bar{q}_a(x) \gamma_5 c_a(x)] \, [\bar{q}_b(x) \gamma^\mu s_b(x)],\\
J_{\mu }^{2}(x) &= [\bar{q}_a(x) \gamma^\mu c_a(x)] \, [\bar{q}_b(x) \gamma_5 s_b(x)],
\label{curr2}\\
J_{\mu }^{3}(x) &=[\bar{q}_a(x) \gamma_\alpha c_a(x)] \, [\bar{q}_b(x) \sigma^{\mu \alpha}\gamma_5 s_b(x)]
                - [\bar{q}_a(x) \sigma^{\mu \alpha}\gamma_5 c_a(x)] \, [\bar{q}_b(x) \gamma_\alpha s_b(x)],
\label{curr3}
\end{align}
where $a$ and $b$ are color indices, and $q$ denotes $u$ or $d$ quark fields.

The current $J_{\mu}^{1}(x)$ in Eq.~(\ref{curr1}) is constructed from the color-singlet bilinears $[\bar{q}\gamma_{5}c]$ and $[\bar{q}\gamma^{\mu}s]$, chosen to match the quantum numbers of a $D\bar{K}^{*}$ system. The first bilinear represents a pseudoscalar charm meson ($D$), while the second corresponds to a vector strange meson ($K^{*}$). In a relative $S$-wave, their product yields total spin-parity $J^{P}=1^{+}$. Therefore, $J_{\mu}^{1}$ predominantly creates a hadronic molecular state with a $D\bar{K}^{*}$ configuration. 
Similarly, the current $J_{\mu}^{2}(x)$ in Eq.~(\ref{curr2}) is built as $[\bar{q}\gamma^{\mu}c] \otimes [\bar{q}\gamma_{5}s]$, corresponding to a vector charm meson ($D^*$) and a pseudoscalar strange meson ($K$). Again, in an $S$-wave the combination gives $J^{P}=1^{+}$, making $J_{\mu}^{2}$ appropriate for a $D^{*}\bar{K}$ molecular configuration. 
The current $J_{\mu}^{3}(x)$ in Eq.~(\ref{curr3}) involves an antisymmetric combination of a vector bilinear and a tensor bilinear,
$
\bigl([\bar{q}\gamma_{\alpha}c][\bar{q}\sigma^{\mu\alpha}\gamma_{5}s] - [\bar{q}\sigma^{\mu\alpha}\gamma_{5}c][\bar{q}\gamma_{\alpha}s]\bigr),
$ 
which is designed to couple to a $D^{*}\bar{K}^{*}$ system in a spin-1 configuration. In a relative $S$-wave between the two vector mesons, this combination yields total spin-parity $J^{P}=1^{+}$, providing the correct Lorentz structure for a molecular interpretation.
The explicit use of color-singlet bilinears reflects the molecular picture of these states: each interpolator is constructed assuming that the physical resonance can be effectively described as a composite of two color-singlet mesons. This is in contrast to diquark–antidiquark currents of the type
$
[q^{T} C c][\bar{s} C \bar{q}^{T}],
$ 
which represents a typical compact tetraquark interpolator and is 
included here for illustrative purposes; such interpolators involve 
color-antitriplet diquarks and antidiquarks and are conceptually 
distinct from the color-singlet bilinear currents used for molecular 
states.

The molecular and compact tetraquark interpretations correspond to 
physically distinct hadronic pictures---loosely bound two-meson 
systems on the one hand, and tightly clustered diquark--antidiquark 
configurations on the other. Within the QCD sum rule framework, 
however, this physical distinction is encoded entirely in the choice 
of interpolating current, while the operator product expansion and 
the matching procedure proceed identically in both cases. 
A formal consequence is that the complete set of independent 
molecular-type currents and the complete set of diquark--antidiquark 
currents are connected by Fierz rearrangement and span the same 
operator basis~\cite{Chen:2021erj, Chen:2010ze, Wang:2020cme, Ozdem:2024rch}. 
A meaningful distinction therefore arises only at the level of a 
single representative current: each individual operator preferentially 
weights a specific Fock component and consequently biases the 
extracted observables toward a particular short-distance configuration. 
The light scalar--isoscalar $\sigma$ meson provides a familiar 
illustration, where a tetraquark-type current---either alone or in 
combination with conventional bilinears---reproduces the properties 
of the state more faithfully than the elementary $\pi\pi$ molecular 
operator. The molecular interpolators in Eqs.~(\ref{curr1})--(\ref{curr3}) 
are adopted because the experimentally observed charm--strange 
tetraquark candidates lie close to their corresponding two-meson 
thresholds, which makes the molecular projection a natural starting 
point. Through the Fierz duality, partial overlap with compact 
configurations is unavoidable, and the electromagnetic moments 
extracted in this work should be regarded as properties of the 
hadronic state populated by these specific currents rather than as 
a model-independent confirmation of the molecular structure.
Using these interpolators, QCD sum-rule calculations can efficiently probe properties that naturally arise from a two-meson structure, including decay constants, production rates, and decay amplitudes~\cite{Chen:2021erj}.

\subsection{QCD description of the correlation function}

Within the framework of QCD, the correlation function is evaluated by inserting the interpolating currents given in Eq.~\eqref{edmn01} and performing all possible contractions using Wick’s theorem. For the hadronic states under consideration, the resulting expressions for the correlation functions are
\begin{align}
T_{\mu\nu}^{\mathrm{QCD}-1}(p,q) &= i\int d^{4}x\, e^{ip\cdot x} \,
\Big\langle 0 \Big\vert 
\mathrm{Tr}\!\Big[ \gamma_{\mu} S_{s}^{bb^{\prime}}(x) \gamma_{\nu} S_{q}^{a^{\prime}a}(-x) \Big]
\mathrm{Tr}\!\Big[ \gamma_{5} S_{c}^{aa^{\prime}}(x) \gamma_{5} S_{q}^{a^{\prime}a}(-x) \Big]
\Big\vert 0 \Big\rangle_{F}, \label{eq:QCDSide1}  
\\
T_{\mu\nu}^{\mathrm{QCD}-2}(p,q) &= i\int d^{4}x\, e^{ip\cdot x} \,
\Big\langle 0 \Big\vert 
\mathrm{Tr}\!\Big[ \gamma_{5} S_{s}^{bb^{\prime}}(x) \gamma_{5} S_{q}^{a^{\prime}a}(-x) \Big]
\mathrm{Tr}\!\Big[ \gamma_{\mu} S_{c}^{aa^{\prime}}(x) \gamma_{\nu} S_{q}^{a^{\prime}a}(-x) \Big]
\Big\vert 0 \Big\rangle_{F}, \label{eq:QCDSide2}
\\
T_{\mu\nu}^{\mathrm{QCD}-3}(p,q) &= i\int d^{4}x\, e^{ip\cdot x} \,
\Big\langle 0 \Big\vert 
\mathrm{Tr}\!\Big[ \sigma_{\mu\alpha}\gamma_{5} S_{s}^{bb^{\prime}}(x) \gamma_{5}\sigma_{\nu\beta} S_{q}^{a^{\prime}a}(-x) \Big]
\mathrm{Tr}\!\Big[ \gamma_{\alpha} S_{c}^{aa^{\prime}}(x) \gamma_{\beta} S_{q}^{a^{\prime}a}(-x) \Big] 
\nonumber\\
& - \mathrm{Tr}\!\Big[ \sigma_{\mu\alpha}\gamma_{5} S_{s}^{bb^{\prime}}(x) \gamma_{\beta} S_{q}^{a^{\prime}a}(-x) \Big]
\mathrm{Tr}\!\Big[ \gamma_{\alpha} S_{c}^{aa^{\prime}}(x) \gamma_{5}\sigma_{\nu\beta} S_{q}^{a^{\prime}a}(-x) \Big] 
\nonumber\\
& - \mathrm{Tr}\!\Big[ \gamma_{\alpha} S_{s}^{bb^{\prime}}(x) \gamma_{5}\sigma_{\nu\beta} S_{q}^{a^{\prime}a}(-x) \Big]
\mathrm{Tr}\!\Big[ \sigma_{\mu\alpha}\gamma_{5} S_{c}^{aa^{\prime}}(x) \gamma_{\beta} S_{q}^{a^{\prime}a}(-x) \Big] \nonumber\\
& + \mathrm{Tr}\!\Big[ \gamma_{\alpha} S_{s}^{bb^{\prime}}(x) \gamma_{\beta} S_{q}^{a^{\prime}a}(-x) \Big]
\mathrm{Tr}\!\Big[ \sigma_{\mu\alpha}\gamma_{5} S_{c}^{aa^{\prime}}(x) \gamma_{5}\sigma_{\nu\beta} S_{q}^{a^{\prime}a}(-x) \Big]
\Big\vert 0 \Big\rangle_{F}, 
\label{eq:QCDSide3}
\end{align}
where the $T_{\mu\nu}^\mathrm{QCD-1}(p,q)$, $T_{\mu\nu}^\mathrm{QCD-2}(p,q)$, and $T_{\mu\nu}^\mathrm{QCD-3}(p,q)$ refer to the molecular configurations $D\bar{K}^{*}$, $D^{*}\bar{K}$, and $D^{*}\bar{K}^{*}$, respectively.

The quark propagators employed in the calculation are taken in coordinate space. For a light quark ($q=u,d ~\mathrm{and}~s$) the propagator reads \cite{Balitsky:1987bk}
\begin{align}
\label{edmn13}
S_{q}(x) &= S_{q}^{\mathrm{free}}(x) 
- \frac{i g_{s}}{16\pi^{2}x^{2}} \int_{0}^{1} du\; G^{\mu\nu}(ux)
\Big[ \bar u \rlap/{x} \sigma_{\mu\nu} + u \sigma_{\mu\nu} \rlap/{x} \Big],
\end{align}
and for a charm quark \cite{Belyaev:1985wza}
\begin{align}
\label{edmn14}
S_{c}(x) &= S_{c}^{\mathrm{free}}(x)
- \frac{i m_{c} g_{s}}{16\pi^{2}} \int_{0}^{1} du\; G^{\mu\nu}(ux)
\Bigg[ (\sigma_{\mu\nu}\rlap/{x} + \rlap/{x}\sigma_{\mu\nu})
\frac{K_{1}\!\big(m_{c}\sqrt{-x^{2}}\big)}{\sqrt{-x^{2}}}
+ 2\sigma_{\mu\nu} K_{0}\!\big(m_{c}\sqrt{-x^{2}}\big) \Bigg].
\end{align}
The free (perturbative) parts are given by
\begin{align}
S_{q}^{\mathrm{free}}(x) &= \frac{1}{2\pi^{2}x^{2}}\Big(i\frac{\rlap/{x}}{x^{2}} - \frac{m_{q}}{2}\Big), \\
S_{c}^{\mathrm{free}}(x) &= \frac{m_{c}^{2}}{4\pi^{2}}
\Bigg[ \frac{K_{1}\!\big(m_{c}\sqrt{-x^{2}}\big)}{\sqrt{-x^{2}}}
+ i\frac{\rlap/{x}\,K_{2}\!\big(m_{c}\sqrt{-x^{2}}\big)}
{(\sqrt{-x^{2}})^{2}} \Bigg],
\end{align}
where $G^{\mu\nu}(x)$ is the gluon field-strength tensor and $K_{n}(z)$ denote modified Bessel functions of the second kind.

The photon can couple to the quark lines either perturbatively, through the elementary QED vertex, or non-perturbatively, via the photon distribution amplitudes (DAs). A consistent treatment requires the inclusion of both mechanisms. Their implementation proceeds as follows:
\begin{itemize}
\item Perturbative (short-distance) photon emission: 
The photon couples directly to a quark line via the substitution
\begin{align}
\label{free}
S_{c(q)}^{\mathrm{free}}(x) \;\longrightarrow\; \int d^{4}z\; S_{c(q)}^{\mathrm{free}}(x-z)\, \rlap/{\!A}(z)\, S_{c(q)}^{\mathrm{free}}(z),
\end{align}
where $A_{\mu}(z)$ is the photon field. This replacement is applied to one quark propagator (light or heavy) while the remaining propagators are kept in their free form.

\item Non-perturbative (long-distance) photon emission:
The photon couples to the QCD vacuum through light-quark condensates and photon DAs. This effect is incorporated by replacing one light-quark propagator according to \cite{Belyaev:1985wza}
\begin{align}
\label{edmn21}
S_{q,\alpha\beta}^{ab}(x) \;\longrightarrow\; -\frac{1}{4}
\big[ \bar{q}^{a}(x) \Gamma_{i} q^{b}(0) \big] (\Gamma_{i})_{\alpha\beta},
\end{align}
with $\Gamma_{i} = \{\mathbb{1},\gamma_{5},\gamma_{\mu},i\gamma_{5}\gamma_{\mu},\sigma_{\mu\nu}/2\}$. The resulting matrix elements of the form $\langle \gamma(q) \vert \bar{q}(x) \Gamma_{i} q(0) \vert 0 \rangle$ and $\langle \gamma(q) \vert \bar{q}(x) $ $ \Gamma_{i} G_{\alpha\beta} q(0) \vert 0 \rangle$ are expressed in terms of photon DAs of increasing twist \cite{Ball:2002ps}. 
In this work, the long-distance (non-perturbative) photon couplings are included only via photon DAs involving light quarks ($u$, $d$, $s$). Contributions from heavy-quark (charm) photon DAs are omitted because they are suppressed by powers of the heavy-quark mass $1/m_c$---a well-established feature in QCD sum rule and effective field theory treatments of heavy hadrons~\cite{Antonov:2012ud}. This suppression reflects the fact that the polarization of the QCD vacuum due to heavy quarks is parametrically smaller than that due to light quarks. Consequently, heavy quarks contribute only through perturbative photon emission, while the non-perturbative photon structure is captured entirely by light-quark degrees of freedom.
\end{itemize}

The systematic combination of the perturbative and non-perturbative photon couplings leads to the complete QCD representation of the correlation function. The explicit evaluation of the resulting integrals, together with a subsequent matching to the hadronic representation, will allow the extraction of the electromagnetic properties of the states under study.

\subsection{Hadronic description of the correlation function}

The hadronic representation of the correlation function is obtained by saturating it with a complete set of hadronic states possessing the same quantum numbers as the interpolating currents. After inserting these intermediate states and isolating the contribution of the lowest-lying $T_{cs}$ tetraquark state, the correlation function can be written as
\begin{align}
\label{edmn04}
T_{\mu\nu}^{\mathrm{Had}}(p,q) &=    \frac{1}{[p^2 - m_{T_{cs}}^2][p^{\prime 2} - m_{T_{cs}}^2] }\langle 0 | J_\mu(x) | T_{cs}(p,\varepsilon^i)\rangle \,
\langle T_{cs}(p,\varepsilon^i) | T_{cs}(p^{\prime},\varepsilon^f)\rangle_{F}\,
\langle T_{cs}(p^{\prime},\varepsilon^f) | J_\nu^\dagger(0) | 0\rangle \nonumber\\
&\quad + \text{higher resonances and continuum contributions}.
\end{align}
To proceed, the explicit forms of the matrix elements appearing in the above expression are required. The overlap between the tetraquark state and the interpolating current is parametrized by
\begin{align}
\label{edmn05}
\langle 0 | J_\mu (x) | T_{cs} (p, \varepsilon^i) \rangle &= \lambda_{T_{cs}} \, \varepsilon_\mu^i, \\
\langle T_{cs} (p^{\prime}, \varepsilon^{f}) | J_{\nu}^{\dagger} (0) | 0 \rangle &= \lambda_{T_{cs}} \, \varepsilon_\nu^{*f},
\end{align}
where $\lambda_{T_{cs}}$ denotes the coupling of the current to the physical state, and $\varepsilon_\mu^i$ ($\varepsilon_\nu^{*f}$) is the polarization vector of the initial (final) tetraquark.

The matrix element describing the interaction of the tetraquark with an external photon, $\langle T_{cs}(p,\varepsilon^i) | T_{cs}(p^{\prime},\varepsilon^f)\rangle_{F}$, can be expressed in terms of three independent Lorentz-invariant form factors $G_1(Q^2)$, $G_2(Q^2)$, and $G_3(Q^2)$ as \cite{Brodsky:1992px}
\begin{align}
\label{edmn06}
\langle T_{cs}(p,\varepsilon^i) | T_{cs}p^{\prime},\varepsilon^f)\rangle_{F} &= - \varepsilon^\gamma (\varepsilon^{i})^\mu (\varepsilon^{f})^\nu
\Big[ G_1(Q^2) (p+p^{\prime})_\gamma \, g_{\mu\nu}
+ G_2(Q^2) ( g_{\gamma\nu}\, q_\mu - g_{\gamma\mu}\, q_\nu ) \nonumber\\
&\quad - \frac{1}{2 m_{T_{cs}}^2} G_3(Q^2) (p+p^{\prime})_\gamma \, q_\mu q_\nu \Big].
\end{align}
Here $\varepsilon^\gamma$ is the polarization vector of the photon, and $Q^2 = -q^2$ is the momentum transfer squared.

In phenomenological analyses, it is customary to employ the magnetic and quadrupole form factors, $F_M(Q^2)$ and $F_{\mathcal{D}}(Q^2)$, which are linear combinations of the $G_i(Q^2)$:
\begin{align}
\label{edmn07}
F_M(Q^2) &= G_2(Q^2), \nonumber \\
F_{\mathcal{D}}(Q^2) &= G_1(Q^2) - G_2(Q^2) + \left(1 + \frac{Q^2}{4m_{T_{cs}}^2}\right) G_3(Q^2).
\end{align}
For real-photon processes ($Q^2 = 0$), these form factors reduce to the static electromagnetic moments of the tetraquark:
\begin{align}
\label{edmn08}
\mu_{T_{cs}} &= \frac{e}{2m_{T_{cs}}}\, F_M(0), \\
\mathcal{D}_{T_{cs}} &= \frac{e}{m_{T_{cs}}^2}\, F_{\mathcal{D}}(0).
\end{align}

Substituting Eqs.~\eqref{edmn05}--\eqref{edmn07} into Eq.~\eqref{edmn04} and expressing the result in terms of $F_M(0)$ and $F_{\mathcal{D}}(0)$ yields the hadronic representation of the correlation function relevant for the static limit:
\begin{align}
T^{\mathrm{Had}}_{\mu\nu}(p,q) &= \frac{\lambda^{2}_{T_{cs}}}{\big[m^{2}_{T_{cs}} - p^{\prime 2}\big]\big[m^{2}_{T_{cs}} - p^{2}\big]}
\Bigg[ F_{M}(0)\Big( q_{\mu}\varepsilon_{\nu} - q_{\nu}\varepsilon_{\mu} + \frac{(\varepsilon\cdot p)}{m^{2}_{T_{cs}}}(p_{\mu}q_{\nu} - p_{\nu}q_{\mu}) \Big)
- F_{\mathcal{D}}(0) \frac{(\varepsilon\cdot p)}{m^{2}_{T_{cs}}} \, q_{\mu}q_{\nu} \Bigg].
\end{align}
The above expression provides the starting point for extracting the magnetic and quadrupole moments of the $T_{cs}$ tetraquark, thereby completing the hadronic side of the analysis.

\subsection{LCSR for the electromagnetic moments}

The extraction of the electromagnetic moments proceeds through a systematic analysis of the correlation function within the LCSR framework. The key step involves matching the hadronic and QCD representations of the correlation function. Specifically, the magnetic moment is obtained by isolating the coefficient of the antisymmetric tensor structure $(\varepsilon_\mu q_\nu - \varepsilon_\nu q_\mu)$, which is unique to the magnetic dipole interaction in the static limit (\(Q^2 = 0\)). After performing the double Borel transformation with respect to the variables \(-p^2\) and \(-p^{\prime 2}\) to suppress continuum contributions, we arrive at the master sum rule
\begin{align}
\mu_{T_{cs}} \, \lambda_{T_{cs}}^2 \; e^{-m_{T_{cs}}^2/\mathrm{M_1^2}} e^{-m_{T_{cs}}^2/\mathrm{M_2^2}}
= T_{\mu\nu}^{\mathrm{QCD}}(\mathrm{M_1^2}, \mathrm{M_2^2}),
\end{align}
where \(T_{\mu\nu}^{\mathrm{QCD}}(\mathrm{M_1^2}, \mathrm{M_2^2})\) denotes the Borel-transformed OPE expression.

To account for the continuum and excited states, the quark–hadron duality is employed. In the double dispersion relation, the physical spectral density is approximated by its OPE counterpart above a duality threshold \(s_0\). For the diagonal correlation function considered here, a natural choice is the triangular duality region
\begin{align}
\frac{s_1}{\mathrm{M_1^2}} + \frac{s_2}{\mathrm{M_2^2}} < \frac{\mathrm{s_0}}{\mathrm{M^2}},
\end{align}
with the effective Borel mass defined as \(\mathrm{M^2} = \mathrm{M_1^2} \mathrm{M_2^2}/(\mathrm{M_1^2} + \mathrm{M_2^2})\). Introducing the variables
\begin{align}
s = \frac{\mathrm{M_1^2} \,s_2 + \mathrm{M_2^2} \,s_1}{\mathrm{M_1^2} + \mathrm{M_2^2}}, \quad u = \frac{\mathrm{M_2^2} \,s_1}{\mathrm{M_1^2} \,s_2 + \mathrm{M_2^2} \,s_1},
\end{align}
the double integral can be reduced to a single one:
\begin{align}
T_{\mu\nu}^{\mathrm{QCD}}(\mathrm{M_1^2}, \mathrm{M_2^2}) = \int_0^{s_0} ds \; e^{-s/\mathrm{M^2}} \; \widetilde{\rho}(s),
\end{align}
where the transformed spectral density \(\widetilde{\rho}(s)\) incorporates an integral over the mixing parameter \(u\).

For elastic scattering, the initial and final tetraquark masses are equal, which allows us to symmetrize the Borel parameters by setting \(\mathrm{M_1^2} = \mathrm{M_2^2} = 2\mathrm{M^2}\). This simplifies the expressions considerably and leads to the final LCSR formulas for the magnetic and quadrupole moments of the three molecular configurations:
\begin{align}
\label{jmu1}
\mu_{D\bar K^*} &= \frac{e^{m_{D\bar K^*}^2/\mathrm{M^2}}}{\lambda_{D\bar K^*}^2} \mathcal{R}_1(\mathrm{M^2}, \mathrm{s_0}), &
\mathcal{D}_{D\bar K^*} &= \frac{e^{m_{D\bar K^*}^2/\mathrm{M^2}}}{\lambda_{D\bar K^*}^2} \mathcal{R}_2(\mathrm{M^2}, \mathrm{s_0}), \\
\label{jmu2}
\mu_{D^*\bar K} &= \frac{e^{m_{D^*\bar K}^2/\mathrm{M^2}}}{\lambda_{D^*\bar K}^2} \mathcal{R}_3(\mathrm{M^2}, \mathrm{s_0}), &
\mathcal{D}_{D^*\bar K} &= \frac{e^{m_{D^*\bar K}^2/\mathrm{M^2}}}{\lambda_{D^*\bar K}^2} \mathcal{R}_4(\mathrm{M^2}, \mathrm{s_0}), \\
\label{jmu3}
\mu_{D^*\bar K^*} &= \frac{e^{m_{D^*\bar K^*}^2/\mathrm{M^2}}}{\lambda_{D^*\bar K^*}^2} \mathcal{R}_5(\mathrm{M^2},\mathrm{s_0}), &
\mathcal{D}_{D^*\bar K^*} &= \frac{e^{m_{D^*\bar K^*}^2/\mathrm{M^2}}}{\lambda_{D^*\bar K^*}^2} \mathcal{R}_6(\mathrm{M^2},\mathrm{s_0}).
\end{align}

The functions \(\mathcal{R}_i(\mathrm{M^2}, \mathrm{s_0})\) (\(i = 1, \dots, 6\)) represent the QCD spectral integrals for each state and moment. The spectral functions \(\mathcal{R}_1(\mathrm{M^2}, \mathrm{s_0})\), \(\mathcal{R}_3(\mathrm{M^2}, \mathrm{s_0})\), and \(\mathcal{R}_5(\mathrm{M^2}, \mathrm{s_0})\) on the right-hand side correspond to the magnetic moments of the \(D\bar{K}^{*}\), \(D^{*}\bar{K}\), and \(D^{*}\bar{K}^{*}\) states, respectively. Their explicit forms, which follow from a lengthy evaluation of the QCD side including both perturbative and non-perturbative photon couplings, are provided in the Appendix for completeness.

\end{widetext}

\section{Numerical Aspects}\label{numerical}

\subsection{Input parameters}

The numerical evaluation of the magnetic and quadrupole moments within the LCSR framework requires the specification of various input parameters, which are collected in Table~\ref{tab:input_parameters}. The photon DAs encode the nonperturbative interaction of the photon with the QCD vacuum and are essential ingredients of the external‑field LCSR. Their explicit forms, which involve the parameter $f_{3\gamma}$ listed in the table, are taken from~\cite{Ball:2002ps}.  The hadron masses $m_{D^{(*)}\bar{K}^{(*)}}$ together with the corresponding current 
couplings $\lambda_{D^{(*)}\bar{K}^{(*)}}$ entering 
Eqs.~(\ref{jmu1})--(\ref{jmu3}) have been determined within the 
same QCD sum rule framework, employing the identical molecular 
interpolating currents in Eqs.~(\ref{curr1})--(\ref{curr3}), in~\cite{Chen:2021erj}, ensuring full internal consistency of 
the present analysis.

\begin{table}[b!]
\centering
\renewcommand{\arraystretch}{1.0}
\caption{Summary of input parameters used in the LCSR analysis. 
The references are: (a)~\cite{ParticleDataGroup:2024cfk}; (b)~\cite{Ioffe:2005ym,Narison:2018nbv}; (c)~\cite{Chen:2021erj}; (d)~\cite{Rohrwild:2007yt}; (e)~\cite{Ball:2002ps}.}
\label{tab:input_parameters}
\begin{tabular}{@{}lcc@{}}
\toprule
Parameter & Value & Ref. \\
\midrule
\textbf{Quark masses} & & \\
\cmidrule(l{2pt}r{2pt}){1-3}
$m_u = m_d$ & $0$ & - \\
$m_s$ & $93.5 \pm 0.08$~MeV & (a) \\
$m_c$ & $1.273 \pm 0.0046$~GeV & (a) \\[0.5em]
\midrule
\textbf{Vacuum condensates} & & \\
\cmidrule(l{2pt}r{2pt}){1-3}
$\langle \bar uu\rangle = \langle \bar dd\rangle$ & $(-0.24 \pm 0.01)^3$~GeV$^3$ & (b) \\
$\langle \bar ss\rangle$ & $(0.8 \pm 0.1)\,\langle \bar uu\rangle$ & (b) \\
$\langle g_s^2 G^2\rangle$ & $0.48 \pm 0.14$~GeV$^4$ & (b) \\[0.5em]

\midrule
\textbf{Tetraquark parameters} & & \\
\cmidrule(l{2pt}r{2pt}){1-3}
$m_{D\bar K^*}$ & $2.89^{+0.10}_{-0.11}$~GeV & (c) \\
$\lambda_{D\bar K^*}$ & $(8.6^{+0.19}_{-0.17}) \times 10^{-2}$~GeV$^5$ & (c) \\
$m_{D^*\bar K}$ & $2.85^{+0.10}_{-0.11}$~GeV & (c) \\
$\lambda_{D^*\bar K}$ & $(0.82^{+0.18}_{-0.17}) \times 10^{-2}$~GeV$^5$ & (c) \\
$m_{D^*\bar K^*}$ & $3.13^{+0.12}_{-0.13}$~GeV & (c) \\
$\lambda_{D^*\bar K^*}$ & $(2.96^{+0.62}_{-0.55}) \times 10^{-2}$~GeV$^5$ & (c) \\[0.5em]

\midrule
\textbf{Photon DAs parameters} & & \\
\cmidrule(l{2pt}r{2pt}){1-3}
$\chi$  & $-2.85 \pm 0.5$~GeV$^{-2}$ & (d) \\
$f_{3\gamma}$  & $-0.0039 $~GeV$^2$ & (e) \\
\bottomrule
\end{tabular}
\end{table}

\subsection{Stability and reliability of LCSR}

According to Eqs.~(\ref{jmu1})--(\ref{jmu3}), the LCSR predictions depend on two auxiliary parameters: the continuum threshold $\mathrm{s_0}$ and the Borel mass parameter $\mathrm{M^2}$. Although these parameters are not physical observables, the extraction of reliable magnetic and quadrupole moments requires identifying regions in which the results exhibit minimal sensitivity to their variation. These regions define the working windows of the sum rules. 
The continuum threshold $\mathrm{s_0}$ accounts for the onset of excited states and continuum contributions in the hadronic representation of the correlation function. Phenomenologically, it is expected to lie slightly above the squared mass of the ground state. Guided by standard QCD sum-rule practice, we adopt the interval
$(m_H + 0.5)^2~\mathrm{GeV}^2 \leq \mathrm{s_0} \leq (m_H + 0.7)^2~\mathrm{GeV}^2$,
where $m_H$ denotes the mass of the hadronic state under consideration. This choice is known to provide stable predictions in closely related analyses and is used throughout this work. 
The Borel window for $\mathrm{M^2}$ is determined by two competing constraints. The lower bound is fixed by the convergence of the OPE, requiring that higher-dimensional condensate contributions remain suppressed. The upper bound is constrained by pole dominance, ensuring that the ground-state contribution is not overwhelmed by the continuum. These requirements are quantified through the conditions

\begin{align}
\text{PC} &=
\frac{\mathcal{R}_i(\mathrm{M^2},\mathrm{s_0})}{\mathcal{R}_i(\mathrm{M^2},\infty)} > 40\%, \qquad
\text{CVG} =
\frac{\mathcal{R}_i^{\mathrm{Dim\,7}}(\mathrm{M^2},\mathrm{s_0})}{\mathcal{R}_i(\mathrm{M^2},\mathrm{s_0})} < 5\%,
\end{align}
where $\mathcal{R}_i^{\mathrm{Dim\,7}}(\mathrm{M^2},\mathrm{s_0})$ denotes the highest-dimensional term retained in the OPE for the invariant function $\mathcal{R}_i(\mathrm{M^2},\mathrm{s_0})$.

  \begin{table}[htb!]
	\addtolength{\tabcolsep}{5pt}
    \caption{Determination of the LCSR parameter windows in $\mathrm{s_0}$ and $\mathrm{M^2}$, showing the resulting PC and CVG  for the electromagnetic observables of the $D^{(\ast)} \bar K^{(\ast)}$ systems.}

	\label{table}
	\begin{ruledtabular}
\begin{tabular}{lccccccc}
	   \\
States & $\mathrm{s_0}\,\,(\mathrm{GeV}^2)$&   $\mathrm{M^2}\,\,(\mathrm{GeV}^2)$& PC\,\,($\%$) & CVG\,\,($\%$)   \\
	   \\
	   \hline\hline
\\
$D   \bar K^\ast $& [11.5, 12.9]&  [1.8, 2.4]& [66.55, 41.07]& $ 0.38$ \\
\\
$D^\ast   \bar K $& [11.5, 12.9]&  [2.0, 2.8]& [68.91, 40.10]& $0.42$  \\
\\
$D^\ast \bar K^\ast $& [13.2, 14.6]&   [2.0, 2.8]&  [70.42, 41.41]& $0.44$ \\
\\
\end{tabular}
\end{ruledtabular}
\end{table}

\begin{figure}[htb!]
\subfloat[]{\includegraphics[width=0.33\textwidth]{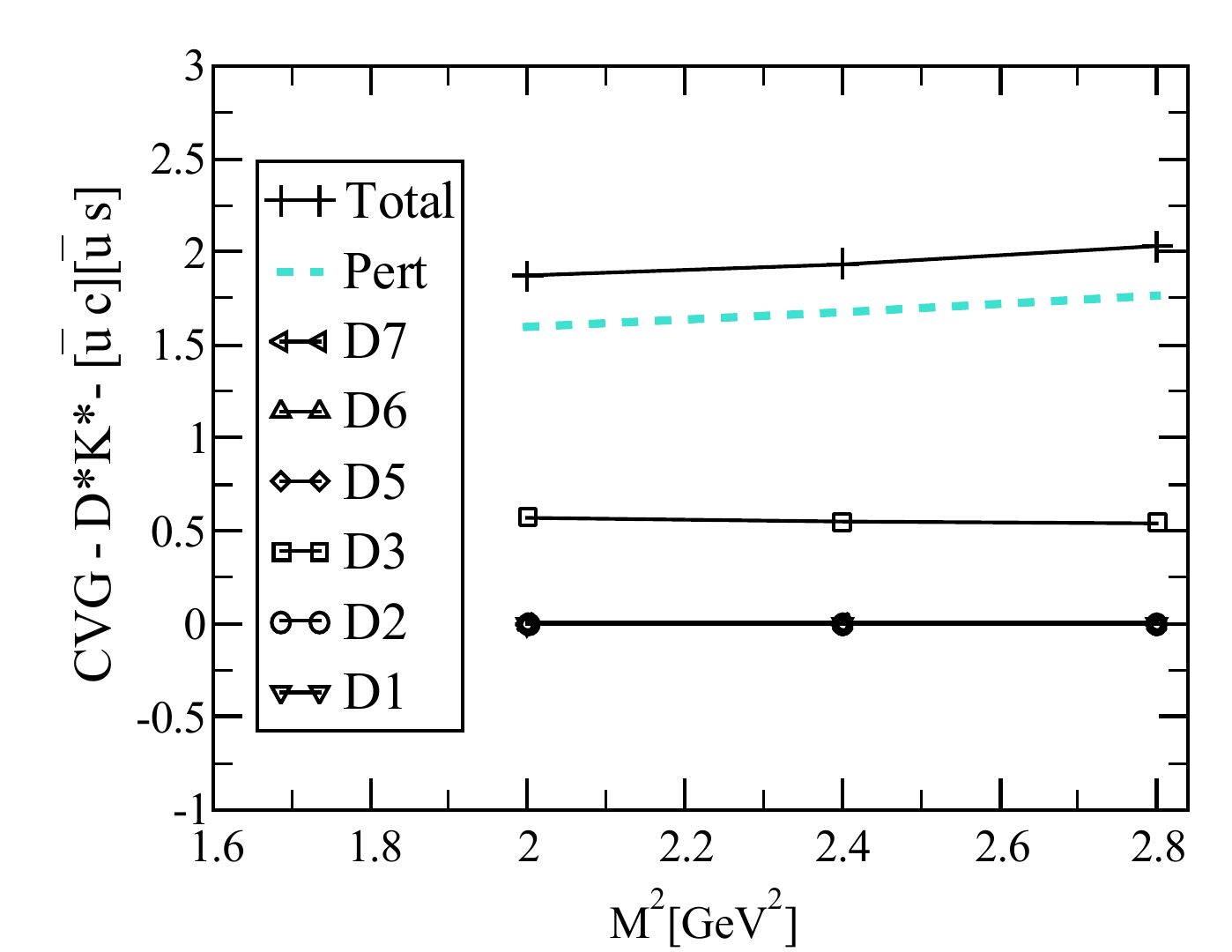}}\qquad \qquad 
\subfloat[]{\includegraphics[width=0.33\textwidth]{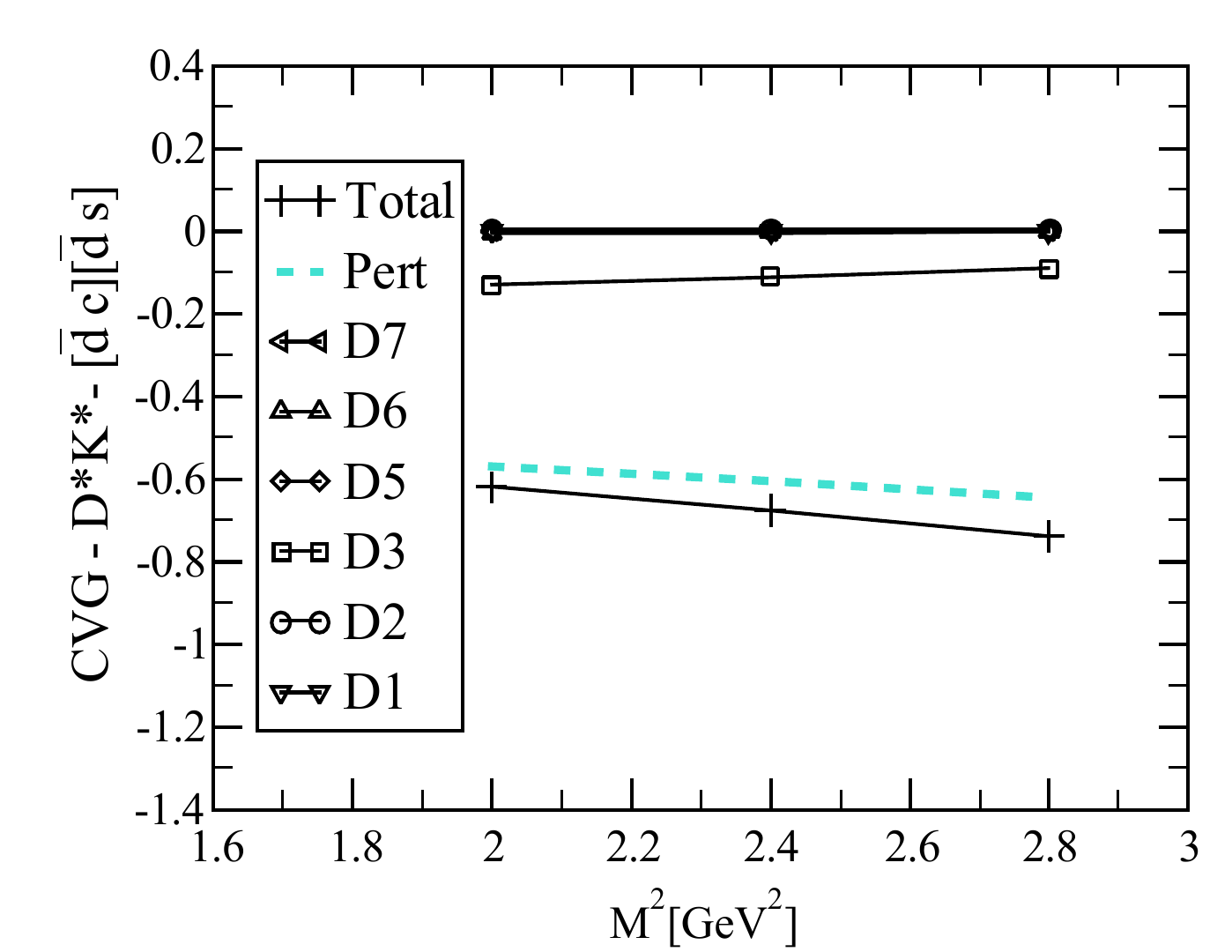}}\\
\subfloat[]{\includegraphics[width=0.33\textwidth]{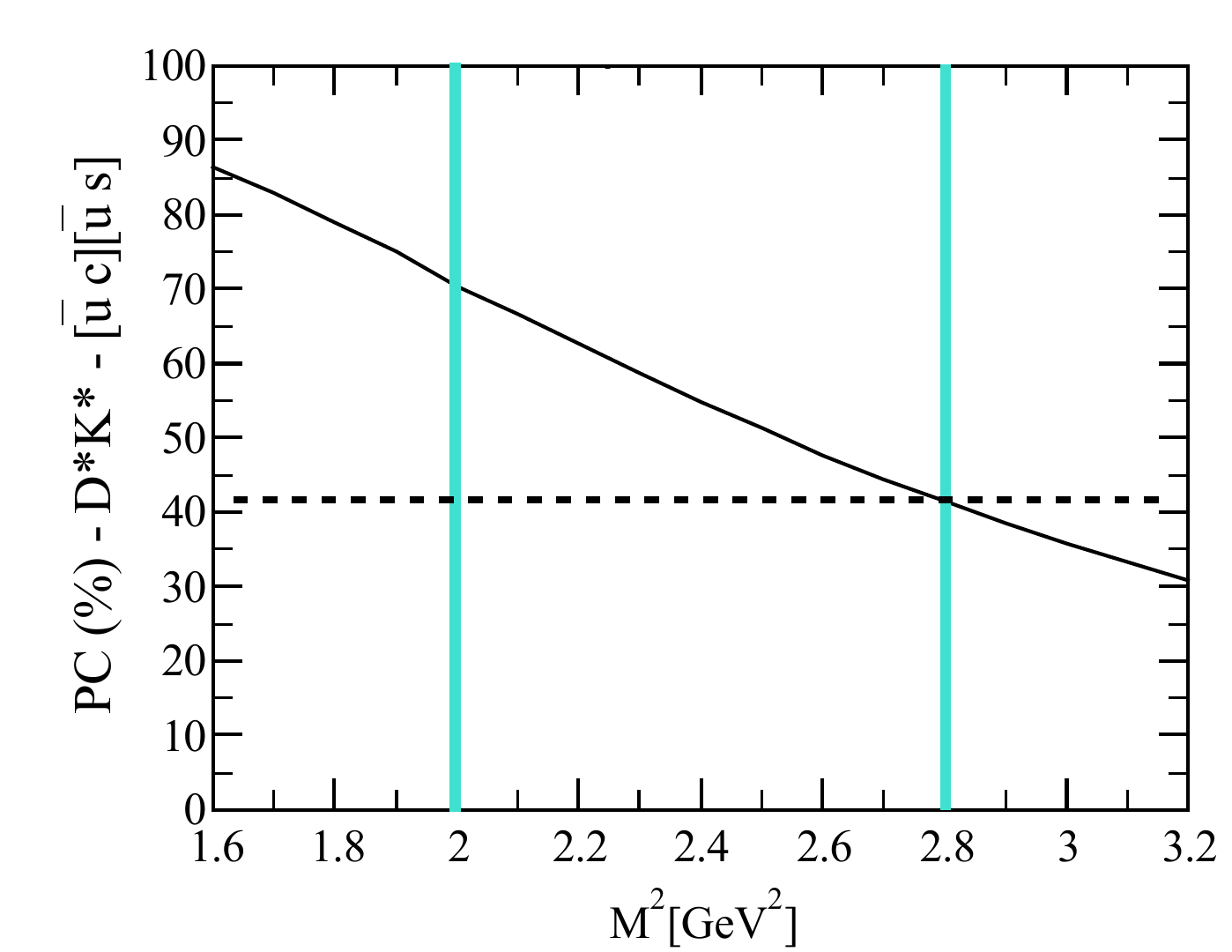}}\qquad \qquad 
\subfloat[]{\includegraphics[width=0.33\textwidth]{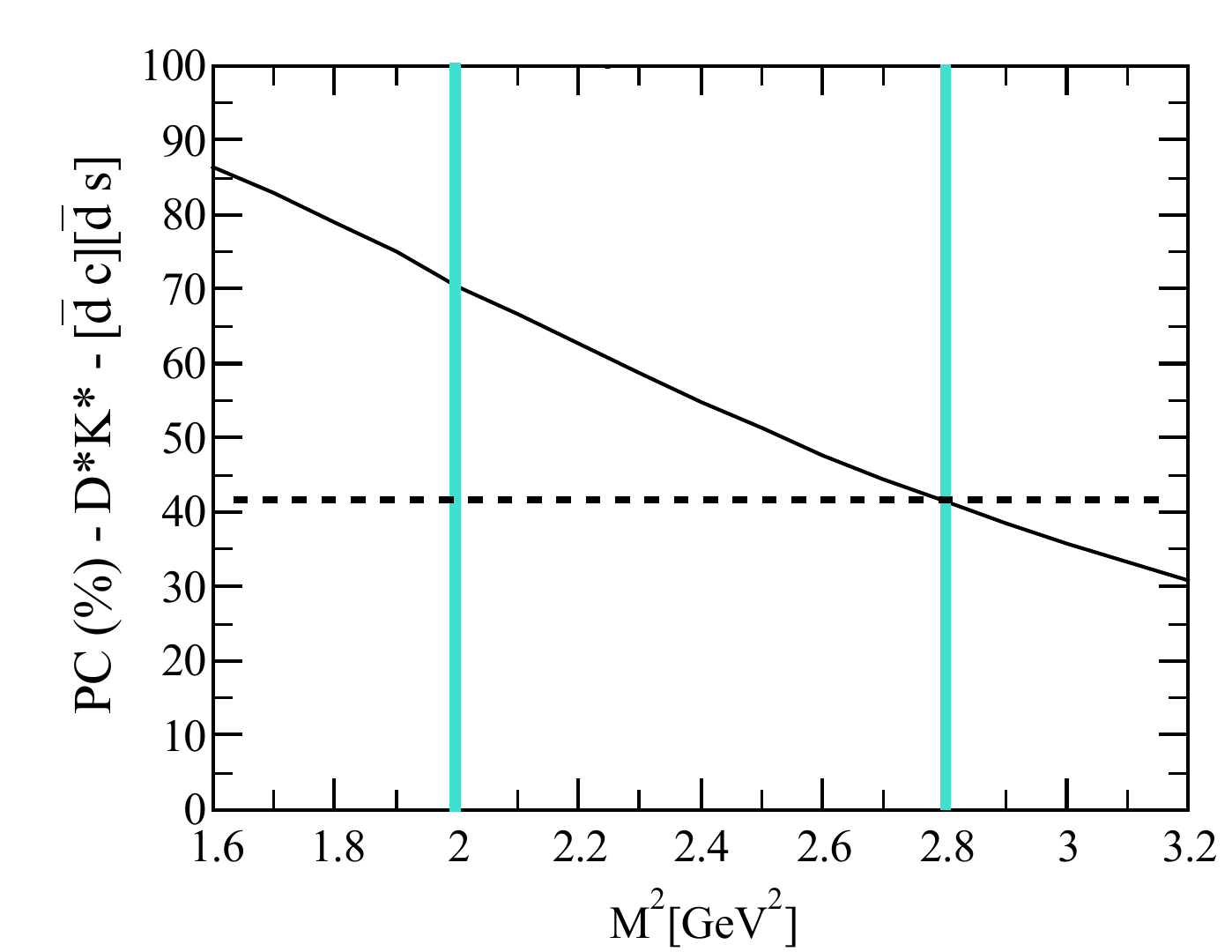}}\\
\subfloat[]{\includegraphics[width=0.33\textwidth]{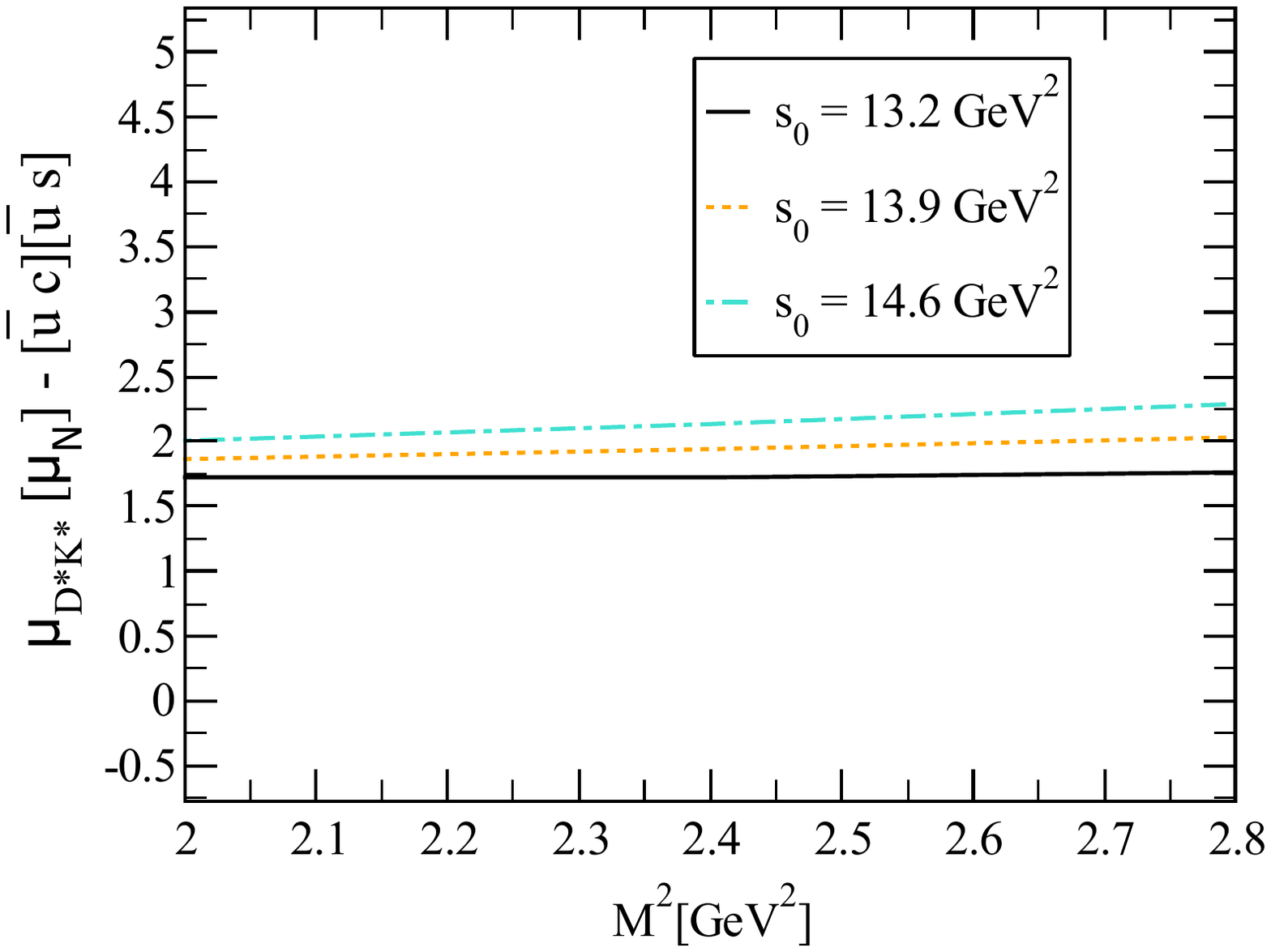}}\qquad \qquad 
\subfloat[]{\includegraphics[width=0.33\textwidth]{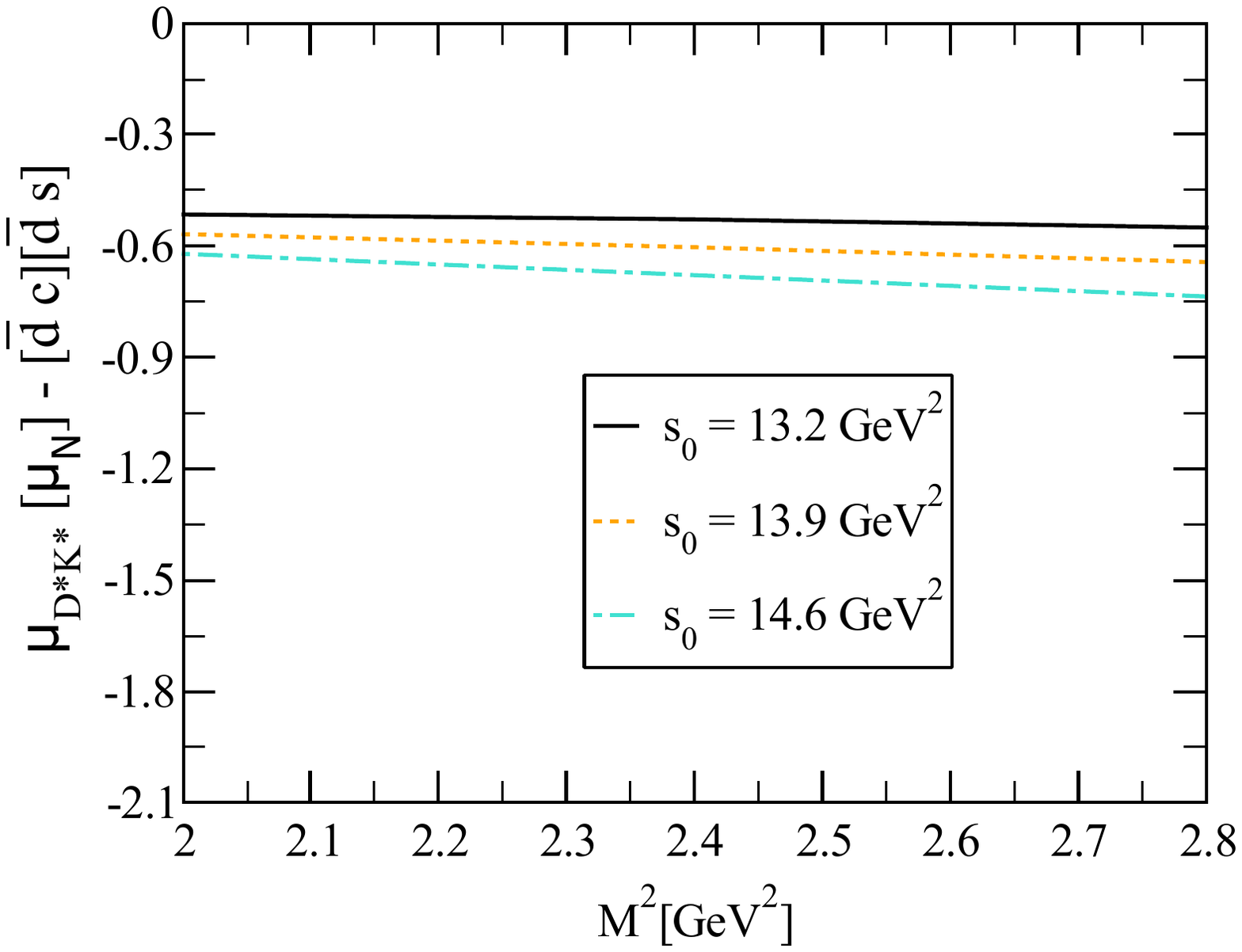}}\\
\caption{Dependence of the sum-rule diagnostics and the extracted magnetic dipole moment on the Borel mass $\mathrm{M^2}$ for the molecular $D^*\bar{K}^*$ configuration. Panels (a) and (b) show the convergence of the CVG for the $[\bar{u}c][\bar{u}s]$ and $[\bar{d}c][\bar{d}s]$ flavor compositions, respectively. Panels (c) and (d) display the corresponding PC. In all four panels, the curves are obtained by varying the continuum threshold $\mathrm{s_0}$ over the interval given in Table~\ref{table}. Panels (e) and (f) show the magnetic moment as a function of $\mathrm{M^2}$ for several fixed representative values of $\mathrm{s_0}$. The vertical dashed lines mark the adopted Borel working window, and the horizontal dashed line in panels (c) and (d) indicates the minimum acceptable pole contribution ($\geq 40\%$).}
\label{Msqfig}
\end{figure}

Imposing these criteria simultaneously, we determine the working regions for $\mathrm{M^2}$ and $\mathrm{s_0}$, summarized in Table~\ref{table}. Within these windows, the PC amounts to $40$--$70\%$ for all channels, confirming the dominance of the lowest-lying state and justifying the single-pole approximation adopted in the hadronic parametrization. At the same time, the OPE exhibits excellent convergence: the relative contribution of the dimension-7 terms remains below $0.5\%$ over the entire Borel window, indicating that the convergence condition is satisfied with a large safety margin well within the adopted $5\%$ criterion.   
The reliability of the sum-rule extraction requires the simultaneous satisfaction of PC and CVG, which together define the fiducial Borel working region.  Within this window, the ground-state contribution must remain sizable compared to the continuum, while higher-dimensional condensate terms stay under control, ensuring the stability of the OPE.  As an illustrative example, Fig.~\ref{Msqfig} displays the CVG, the PC, and the resulting magnetic moment as functions of the Borel parameter $\mathrm{M^2}$ for several fixed values of $\mathrm{s_0}$ in the $D^\ast \bar K^\ast$ channel.  One observes that the PC dominates over the continuum and the OPE remains well convergent inside the adopted Borel window, while the extracted magnetic moment exhibits only a mild dependence on the auxiliary parameters $\mathrm{M^2}$ and $\mathrm{s_0}$. These residual variations are incorporated into the quoted theoretical uncertainties. The fulfillment of the standard LCSR stability criteria suggests that the extracted predictions are reasonably reliable within the present analysis.

A critical assumption underlying Eq.~(\ref{edmn04}) is that the hadronic side of the LCSR  can be adequately represented by a single-pole contribution. For conventional hadrons such as the nucleon or $\phi$ meson, this approximation is well justified. However, for multiquark exotic states, the physical spectral density also 
receives contributions from intermediate two-meson scattering states, as discussed in~\cite{Weinberg:2013cfa, Lucha:2021mwx, Kondo:2004cr, Lucha:2019pmp}. A careful treatment of these effects is therefore necessary when extracting the properties of multiquark candidates. 
Two complementary strategies have been developed in the literature to handle such contributions. The first approach, commonly applied to pentaquark systems~\cite{Sarac:2005fn, Wang:2019hyc, Lee:2004xk}, consists of explicitly subtracting the two-meson continuum from the sum rules. The second method, employed extensively for tetraquark studies~\cite{Wang:2020cme, Albuquerque:2020hio, Albuquerque:2021tqd,   Sundu:2018nxt, Wang:2015nwa, Agaev:2018vag,  Wang:2020iqt, Wang:2019igl}, incorporates the effect of the two-meson intermediate states into an effective width of the pole term. In the latter case, the quark propagator is modified as
\begin{equation}
\frac{1}{m^{2}-p^{2}} \longrightarrow \frac{1}{m^{2}-p^{2}-i\sqrt{p^{2}}\,\Gamma(p)},
\label{eq:Modif}
\end{equation}
where $\Gamma(p)$ denotes the energy-dependent decay width generated by the coupling to two-meson channels. Systematic analyses within this framework have demonstrated that the inclusion of such effects leads to corrections in the range of $5$--$7\%$ in the extracted 
physical parameters~\cite{Wang:2015nwa, Agaev:2018vag, Sundu:2018nxt,Albuquerque:2021tqd, Albuquerque:2020hio, Wang:2020iqt, Wang:2019igl, Wang:2020cme}. These corrections remain well within the typical uncertainties of QCD sum-rule calculations, which in the present work amount to approximately $25$--$30\%$, and are therefore negligible at the level of precision of the present analysis.

Furthermore, we emphasise that the pole contribution in our analysis lies in the range of $40$--$70\%$ across all channels (see Table~\ref{table}), confirming that the ground-state pole is not overwhelmed by the two-meson continuum within the adopted Borel 
window. This level of pole dominance is consistent with the standard reliability criteria of QCD sum rules and demonstrates that the single-pole approximation is not undermined by the S-wave $D^{(*)}\bar{K}^{(*)}$ scattering states in the spectral density. We further note that the Borel transformation, combined with the choice of the continuum threshold $s_0$, helps to suppress contributions from higher-lying continuum states. Since the electromagnetic moments are extracted from spin-dependent Lorentz structures that do not 
overlap with the scalar/pseudoscalar quantum numbers of the S-wave two-meson continuum, the contamination from scattering states is expected to be further reduced compared to the mass sum rules discussed in the literature. We therefore consider the zero-width single-pole approximation employed in Eq.~(\ref{edmn04}) to be well justified within the precision of the present study, while acknowledging that a fully coupled-channel treatment incorporating the two-meson continuum explicitly would be a desirable refinement in future, more dedicated analyses.

\subsection{Numerical results}

A comprehensive numerical analysis is performed to extract the magnetic and electric quadrupole moments of the $D^{(\ast)}\bar K^{(\ast)}$ molecular tetraquark states. The resulting central values and uncertainties are summarized in re summarized in Tables~\ref{table} and~\ref{table2}. The quoted errors include the systematic propagation of all input parameters as well as the variations of the auxiliary LCSR parameters, $\mathrm{M^2}$ and $\mathrm{s_0}$. The relative contributions to the total uncertainty are estimated to be approximately $15\%$ from the tetraquark masses, 10\% quark masses, $22\%$ from the residue parameters, $30\%$ from the continuum threshold $\mathrm{s_0}$, $8\%$ from the Borel mass parameter $\mathrm{M^2}$, $10\%$ from the photon DAs, and about $5\%$ from remaining sources such as QCD condensates. All individual uncertainties are combined in quadrature to obtain the final error bars reported in Table~\ref{table2}.


  \begin{table}[htb!]
	\addtolength{\tabcolsep}{5pt}
\caption{Magnetic dipole  and electric quadrupole moments of the $D^{(*)}\bar K^{(*)}$ molecular states with quark contents $[\bar q c][\bar q s]$ $(q=u,d)$ obtained from LCSR.}
	\label{table2}
	\begin{ruledtabular}
\begin{tabular}{lccccccc}
	   \\
 \qquad States \qquad\qquad & \qquad \qquad$\mu\, (\mu_N)$\qquad \qquad&  \qquad \qquad$\mathcal{D}~(\times 10^{-3}\, \mbox{fm}^{2})$  \\
	   \\
	   \hline\hline
	  \\
$D   \bar K^\ast - [\bar u c][\bar u s]$ \qquad& \qquad $~~ 1. 99 \pm 0.49$ \qquad& \qquad $~~3.60 \pm 0.40$ 
\\
\\
$D   \bar K^\ast - [\bar d c][\bar d s]$ \qquad& \qquad $~~ 0. 00 \pm 0.00$ \qquad& \qquad ~~$0.00 \pm 0.00$ 
\\
\\
\hline
\\
$D^\ast   \bar K - [\bar u c][\bar u s]$\qquad& \qquad $~~3.08 \pm 0.77$ \qquad& \qquad $~~6.01 \pm 0.80$ \\
\\
$D^\ast \bar K- [\bar d c][\bar d s]  $ \qquad& \qquad $-2.04 \pm 0.50$ \qquad& \qquad $-3.01 \pm 0.40$
\\
\\
\hline
\\
$D^\ast \bar K^\ast- [\bar u c][\bar u s]  $ \qquad& \qquad $~~1.93\pm 0.47$ \qquad& \qquad $-0.46 \pm 0.08$\\
\\
$D^\ast \bar K^\ast- [\bar d c][\bar d s]  $ \qquad& \qquad $-0.62\pm 0.15$ \qquad& \qquad $-0.09 \pm 0.01$\\
\\
\end{tabular}
\end{ruledtabular}
\end{table}

Our analysis provides quantitative information on the electromagnetic structure of the  $D^{(*)} \bar K^{(*)}$ molecular configurations. In the following, we summarize the main features of the magnetic and quadrupole moments and discuss their physical implications.

\subsubsection{Magnetic moments}

The extracted magnetic moments display a clear and systematic hierarchy among the $D^{(*)}\bar K^{(*)}$ molecular configurations. The largest value is obtained for the charged $D^*\bar K$ state, $\mu = 3.08 \pm 0.77\,\mu_N$, whereas the $D\bar K^*$ and $D^*\bar K^*$ channels yield smaller but comparable magnitudes of about $2\,\mu_N$. This behavior can be traced to the different spin structures of the constituent mesons: the vector--pseudoscalar configuration allows a more efficient alignment of the light-quark spins with the external electromagnetic field, while in the vector--vector system partial cancellations reduce the net magnetic response. A qualitatively different pattern is observed for neutral configurations. The $D\bar K^*$ state exhibits a vanishing magnetic moment as a consequence of charge cancellation combined with the specific spin--flavor structure of the current, whereas the neutral $D^*\bar K$ and $D^*\bar K^*$ states remain nonzero. This indicates that these molecular combinations are not eigenstates of charge conjugation and retain nontrivial internal spin distributions capable of generating a finite magnetic moment.

The flavor decomposition (see Table~\ref{table3}) further demonstrates a clear dominance of the light quarks. In all channels the $u(d)$ components provide the leading contribution to the total magnetic moment. The strange-quark term remains subleading but non-negligible, typically $|\mu_s|\sim 0.4$–$0.7\,\mu_N$, reflecting the intermediate mass of the $s$ quark relative to the light sector. By contrast, the charm-quark contribution is strongly suppressed, $|\mu_c|\lesssim 0.3\,\mu_N$, in accordance with heavy-quark symmetry, where magnetic moments scale approximately as $1/m_c$. These results establish the hierarchy
\begin{equation*}
|\mu_{u(d)}| > |\mu_s| > |\mu_c|,
\end{equation*}
showing that the electromagnetic response is governed predominantly by the light constituents, while the heavy quark acts essentially as a static color source.

Consequently, at the hadronic level one obtains the ordering
\begin{equation*}
\mu_{D^*\bar K} > \mu_{D\bar K^*} \simeq \mu_{D^*\bar K^*},
\end{equation*}
which is qualitatively consistent with expectations for a loosely bound molecular configuration, where the magnetic response is dominated by the light-quark degrees of freedom.

  \begin{table}[htb!]
	\addtolength{\tabcolsep}{6pt}
	\caption{Individual quark-flavor contributions to the magnetic moments of the $D^{(\ast)} \bar K^{(\ast)}$ molecular states. All quantities are expressed in units of $\mu_{N}$. Here, the total magnetic moment is given by 
$\mu_{\mathrm{tot}} = \mu_{u(d)} + \mu_s + \mu_c$, 
corresponding to the sum of the individual quark contributions. }
	\label{table3}
	\begin{ruledtabular}
\begin{tabular}{lccccccc}
	   \\
States &   $\mu_{u}$& $\mu_d $ & $\mu_s$& $\mu_{c}$  &$\mu_{tot} $\\
	   \\
	   \hline\hline
\\
$D   \bar K^\ast - [\bar u c][\bar u s]$&  $~~1.33$& $-$& $~~0.66$&$~~0.00$ &$~~1.99$\\
\\
$D   \bar K^\ast - [\bar d c][\bar d s]$&  $-$& $-0.66$& $~~0.66$&$~~0.00$ &$~~0.00$\\
\hline
\\
$D^\ast   \bar K - [\bar u c][\bar u s]$&  $~~3.41$& $-$& $~~0.00$&$-0.33$ &$~~3.08$\\
\\
$D^\ast   \bar K - [\bar d c][\bar d s]$&  $-$& $-1.71$& $~~0.00$&$-0.33$ &$-2.04$\\
\hline
\\
$D^\ast   \bar K^\ast - [\bar u c][\bar u s]$&  $~~1.70$& $-$& $~~0.44$&$-0.21$ &$~~1.93$\\
\\
$D^\ast   \bar K^\ast - [\bar d c][\bar d s]$&  $-$& $-0.85$& $~~0.44$&$-0.21$ &$-0.62$\\

\end{tabular}
\end{ruledtabular}
\end{table}

\subsubsection{Quadrupole moments}

The electric quadrupole moments probe deviations from spherical charge
distributions and therefore provide complementary information to the
magnetic dipole moments about the internal spatial structure of the
states. In contrast to the dipole moments, which are controlled mainly by the total spin and flavor content, the quadrupole moments are sensitive to the anisotropy of the charge distribution and require nonzero orbital or tensor components in the wave function. Numerically, all obtained values are of order $10^{-3}\,\mathrm{fm}^2$ and are significantly suppressed relative to the magnetic moments. Since a quadrupole moment scales parametrically as $\mathcal D \sim e r^2$, its small magnitude suggests only weak deviations from spherical symmetry.

Among the considered systems, the $D^*\bar K$ states exhibit the largest absolute magnitudes, whereas the fully neutral $D\bar K^*$ configuration vanishes identically due to charge symmetry. Positive values correspond to prolate shapes and negative ones to oblate deformations; accordingly, the $[\bar u c][\bar u s]$ channel is prolate, while the $[\bar d c][\bar d s]$ and both $D^*\bar K^*$ configurations are oblate. Although the dominant configuration of these molecular candidates is expected to be $S$ wave, relativistic quark–gluon dynamics and spin–orbit correlations within the  LCSR framework can be interpreted as arising from small $D$-wave or tensor admixtures, which in turn generate finite quadrupole moments. The resulting hierarchy,
\begin{equation*}
|\mathcal{D}_{D^*\bar K}| >
|\mathcal{D}_{D\bar K^*}| >
|\mathcal{D}_{D^*\bar K^*}|,
\end{equation*}
suggests that such anisotropic components are most pronounced in the
$D^*\bar K$ channel, likely reflecting a stronger interplay between the vector meson polarization and the spatial separation of the light-quark charges.

Overall, the modest magnitude of the quadrupole moments indicates that
these states are only weakly deformed and spatially extended, which is qualitatively consistent with expectations for loosely bound
molecular configurations, although a definitive discrimination from more compact multiquark scenarios would require additional observables.

\subsubsection{Structural fingerprints and discrimination from compact 
configurations} \label{subsec:structure}

A natural concern in any LCSR analysis of multiquark states is whether the extracted electromagnetic moments genuinely reflect the molecular configuration adopted in the interpolating currents, or whether alternative structural assignments---most notably compact 
diquark--antidiquark tetraquarks---would yield comparable predictions. As discussed in Sec.~\ref{sec:IIB}, the Fierz duality between molecular and diquark--antidiquark operators implies that no single observable can provide a model-independent confirmation of the molecular picture. Nevertheless, the present results identify several qualitative features that, taken together, constitute structural fingerprints of the molecular scenario rather than predictions of the moment hierarchy alone.

\medskip
\noindent\textit{(i) Vanishing magnetic and quadrupole moments of the 
neutral $D\bar{K}^{*}$:}
The $[\bar{d}c][\bar{d}s]$ configuration of $D\bar{K}^{*}$ yields 
$\mu = 0$ and $\mathcal{D} = 0$ identically 
(Tables~\ref{table2},~\ref{table3}). Two distinct mechanisms are 
at work in this exact cancellation. First, the light-quark 
contributions $\mu_d$ from the $[\bar{d}c]$ and $[\bar{d}s]$ clusters 
combine with opposite signs and equal magnitudes, reflecting the 
overall electric neutrality of the $D^{0}\bar{K}^{*0}$ system. 
Second, and more remarkably, the charm-quark contribution $\mu_c$ 
vanishes analytically: a careful inspection of the QCD-side expression for $\mathcal{R}_1(\mathrm{M^2}, \mathrm{s_0})$ given in the Appendix reveals that no term proportional to $e_c$ survives for this molecular current, a feature dictated by the spin-Lorentz structure of the bilinear-bilinear operator $[\bar{q}\gamma_5 c][\bar{q}\gamma^\mu s]$. This analytic suppression is a structural fingerprint of the molecular interpolator: in a compact diquark--antidiquark current, where the charm quark is correlated with a light quark within a color-antitriplet cluster of different Lorentz structure, no such cancellation is generically protected, and a nonzero $\mu_c$ would be expected. The simultaneous vanishing of $\mu_c$ analytically and of $\mu_{u(d)} + \mu_s$ via charge balance therefore constitutes a sharp, multi-layered signature of the two-meson assignment.

\medskip
\noindent\textit{(ii) Flavor hierarchy and charm-quark suppression:}
The flavor decomposition in Table~\ref{table3} exhibits a strong 
dominance of light-quark contributions, with 
$|\mu_{u(d)}| > |\mu_{s}| > |\mu_{c}|$ across all channels. The 
charm-quark contribution remains bounded by 
$|\mu_{c}| \lesssim 0.3\,\mu_{N}$, in line with the heavy-quark 
scaling $\mu_{c} \sim 1/m_{c}$. Within the molecular picture this 
suppression is natural: the charm quark is localized inside the 
$D^{(*)}$ meson and contributes only through perturbative photon 
emission, while the long-distance photon structure is carried entirely by the light degrees of freedom. Compact diquark--antidiquark configurations, in which the heavy quark participates in a tightly correlated spin-1 cluster with a light quark, generically yield substantially larger magnetic responses. This expectation is borne out by an explicit parallel LCSR analysis of the closely related open-flavor $J^{P}=1^{+}$ $B_{c}$-like tetraquarks, where 
diquark--antidiquark interpolators were found to produce magnetic 
moments a factor of $\sim 2.5$--$7$ larger than the corresponding 
molecular ones~\cite{Ozdem:2022eds}. A particularly compelling 
illustration of this structural sensitivity is provided by the 
$Z_{c}(3900)$, for which the same physical state has been analyzed within both the diquark--antidiquark picture using LCSR~\cite{Ozdem:2017jqh} and the molecular picture using QCD sum rules in the external electromagnetic field~\cite{Xu:2020qtg}. The two interpretations yield magnetic moments that differ by more than a factor of three ($\mu_{Z_c} \simeq 0.67\,\mu_{N}$ in the diquark--antidiquark picture versus $\mu_{Z_c} \simeq 0.19\,\mu_{N}$ in the molecular one), demonstrating that the magnetic dipole moment can resolve structural assignments even when the underlying physical state is common to both pictures. A similar sensitivity would be expected for compact $D^{(*)}\bar{K}^{(*)}$ configurations, so that a sizable charm-quark contribution to the total magnetic moment would signal a more compact internal arrangement than the molecular assignment adopted here. 

\medskip
\noindent\textit{(iii) Smallness of the quadrupole deformations:}
All extracted quadrupole moments are of order 
$10^{-3}\,\mathrm{fm}^{2}$, indicating only weak deviations from 
spherical charge distributions. Since $\mathcal{D} \sim e\langle 
r^{2}\rangle$ scales with the spatial anisotropy of the charge 
distribution, the present values point to nearly spherical 
configurations with only weak tensor admixture, qualitatively 
consistent with a loosely bound, spatially extended two-meson system 
in which the dominant $S$-wave component carries no intrinsic 
deformation. Compact diquark--antidiquark configurations, 
characterized by stronger tensor correlations and shorter internal 
distances, would generically support larger quadrupole deformations: 
in the analogous open-flavor $J^{P}=1^{+}$ $B_{c}$-like tetraquark 
analysis of~\cite{Ozdem:2022eds}, the diquark--antidiquark 
quadrupole moments were found to exceed the molecular ones by 
factors of about $3$--$4$, in line with this expectation.

\medskip
\noindent\textit{(iv) Reading the moment hierarchy:}
Within the above context, the ordering 
$\mu_{D^{*}\bar{K}} > \mu_{D\bar{K}^{*}} \simeq \mu_{D^{*}\bar{K}^{*}}$ should be interpreted not as a structural discriminant in isolation, but as one element of a broader set of qualitative signatures. The hierarchy itself is largely controlled by light-quark charges and spin alignments and is therefore expected to be present, with similar qualitative ordering, in a range of structural assignments. What distinguishes the molecular scenario in our analysis is the combination of the moment hierarchy with the vanishing neutral $D\bar{K}^{*}$ moments, the charm-quark suppression, and the small quadrupole deformations---a pattern for which the closely related $B_{c}$-like analysis of~\cite{Ozdem:2022eds} provides direct quantitative support of the expected molecular-vs-compact difference.


\subsubsection{Phenomenological consequences and accessible observables}
\label{subsec:phenomenology}

Although the magnetic and quadrupole moments are static quantities, they encode information about the electromagnetic structure of the states and may affect processes involving real or virtual photons. Consequently, they can lead to potentially observable effects in production mechanisms, radiative transitions, and other electromagnetic probes.

States with comparatively large magnetic moments are expected to exhibit enhanced couplings to external electromagnetic fields and may therefore be more sensitive to photon-induced production channels. At the level of dimensional analysis, or within simple effective hadronic descriptions, the near-threshold photoproduction cross section for reactions such as $\gamma p \to T_{cs}+X$ can be parametrically estimated to scale schematically as
\begin{equation*}
\sigma(\gamma p \to T_{cs}+X)
\propto
\frac{\alpha}{m_{T_{cs}}^{2}}\,\mu_{T_{cs}}^{2},
\end{equation*}
which should be interpreted only as an order-of-magnitude estimate rather than a quantitative prediction. In practice, realistic cross sections depend on additional ingredients such as hadronic form factors, production dynamics, and final-state interactions, effects that lie beyond the scope of the present LCSR framework. Nevertheless, magnetic moments of order $\mu_{T_{cs}}\sim 2\!-\!3\,\mu_N$, as obtained here, suggest that electromagnetic production mechanisms may be experimentally accessible. Similar qualitative sensitivities may also arise in initial-state radiation processes such as $e^+e^- \to T_{cs}\gamma$ or $pp \to T_{cs}\gamma+X$, where the magnetic moment can influence both the overall rate and the angular distributions of the emitted photon.

Electromagnetic moments also enter radiative transitions between possible partner states. If excited configurations exist, the widths of $T_{cs}^* \to T_{cs}\gamma$ decays are governed by magnetic dipole (M1) matrix elements closely related to the static moments and scale approximately as
\begin{equation*}
\Gamma(T_{cs}^* \to T_{cs}\gamma)
\sim
\frac{\alpha}{3}\,\frac{k^3}{m_{T_{cs}}^{2}}\, |\mu_{T_{cs}}|^{2},
\end{equation*}
where $k$ denotes the photon momentum in the rest frame. The same magnetic form factor also controls Dalitz decays through $F_M(Q^2)$, providing additional indirect access to the electromagnetic structure. Furthermore, strong external electromagnetic fields generated in heavy-ion or polarized-hadron environments may induce spin-alignment effects for states with nonzero magnetic moments, offering complementary, though experimentally challenging, probes of their internal dynamics.

The phenomenological consequences discussed above are static-moment 
manifestations of the structural fingerprints analyzed in 
Sec.~\ref{subsec:structure}: enhanced photoproduction rates for states with larger magnetic moments, distinct M1 transition widths between partner states, and possible spin-alignment effects in strong-field environments all derive from the same underlying flavor and spin structure that distinguishes the molecular scenario from compact 
diquark--antidiquark configurations. Future high-luminosity 
experiments at LHCb, Belle II, and BESIII, together with lattice 
QCD calculations and dedicated effective-field-theory studies of 
heavy-meson molecules, will be essential for confronting these 
predictions and clarifying the internal structure of the 
$D^{(\ast)}\bar K^{(\ast)}$ family.

\section{Discussion and Outlook}
\label{sum}

In this work, we have investigated the electromagnetic properties of three charm--strange molecular configurations, $D\bar K^*$, $D^*\bar K$, and $D^*\bar K^*$, with quantum numbers $J^P=1^+$. Using the framework of QCD light--cone sum rules, analytic expressions for the magnetic and quadrupole moments were obtained by matching the hadronic and operator--product--expansion representations of an appropriate correlation function. Both short--distance photon couplings and long--distance contributions encoded in photon distribution amplitudes were consistently taken into account. After fixing the auxiliary parameters according to standard sum--rule criteria, numerical results were extracted and summarized in Tables~\ref{table2} and~\ref{table3}. The reliability of the sum--rule analysis is supported by standard consistency checks. In the selected Borel windows, pole contributions lie in the range of $40\%$--$70\%$, and the convergence of the operator--product expansion is excellent. In particular, the relative contribution of the highest--dimensional (dimension--7) terms remains below $0.5\%$ for all channels, demonstrating that the truncation of the OPE is well under control. The quoted uncertainties arise mainly from the variations of the continuum threshold, hadron masses and residues, the Borel parameter, and the nonperturbative input parameters.

An analysis of flavor contributions shows that the magnetic response is largely governed by the light quarks, while the charm--quark contribution is suppressed by the heavy--quark mass. This behavior is naturally compatible with the interpretation of these systems as loosely bound hadronic molecules, where the heavy quark acts predominantly as a static color source. The fully neutral $D\bar K^*$ configuration has vanishing electromagnetic moments due to overall charge neutrality, whereas the neutral $D^*\bar K$ and $D^*\bar K^*$ states acquire nonzero moments since they are not charge--conjugation eigenstates in the molecular picture.

To our knowledge, this work represents one of the first dedicated 
QCD light--cone sum rule studies of electromagnetic moments for 
$D^{(\ast)}\bar K^{(\ast)}$ molecular states. The results provide 
quantitative benchmarks that can be compared with future predictions 
from constituent quark models, effective field theories for 
heavy--meson molecules, or lattice QCD calculations. Beyond the 
moment hierarchy itself, our analysis identifies several qualitative 
structural fingerprints of the molecular scenario: the exact 
vanishing of the electromagnetic moments of the neutral 
$D\bar K^{*}$ (traceable to an analytic suppression of the 
charm-quark contribution dictated by the spin-Lorentz structure of 
the molecular current), the strong charm-quark suppression in the 
flavor decomposition, and the smallness of the quadrupole 
deformations. Taken together with the parallel molecular-vs-compact 
LCSR analysis of the related $B_{c}$-like tetraquark 
system~\cite{Ozdem:2022eds}---where diquark--antidiquark 
interpolators yield magnetic and quadrupole moments several times 
larger than the molecular ones---and with the $Z_{c}(3900)$ case, 
in which the same physical state analyzed within molecular and 
diquark--antidiquark assignments yields magnetic moments differing 
by more than a factor of three~\cite{Ozdem:2017jqh, Xu:2020qtg}, 
these features suggest that electromagnetic moments offer a useful 
structural diagnostic complementary to mass spectra and decay 
widths, and may help discriminate between molecular configurations 
and more compact multiquark interpretations.

From the perspective of the phenomenology of open-flavor charm--strange exotic states reported by LHCb, several candidates with quantum numbers $J^P=0^+$ and $1^-$ are frequently discussed in the literature as possible $D^{(\ast)}\bar K^{(\ast)}$ hadronic molecules. Although the present analysis focuses on the $D^{(\ast)}\bar K^{(\ast)}$ configuration with $J^P=1^+$, the electromagnetic properties obtained here may still provide qualitative insight into the internal structure of nearby molecular states in the same mass region.

In conclusion, electromagnetic moments provide valuable insight into the internal structure of exotic hadrons. The present analysis indicates that charm--strange molecular states exhibit characteristic magnetic and quadrupole moments that are compatible with an extended two--meson configuration. We expect that ongoing experimental efforts, together with further theoretical developments, will help clarify the nature of these states and deepen our understanding of multiquark dynamics.


\appendix 

\section*{Appendix: Sum rules for the magnetic dipole moments }

The resulting sum rules for the magnetic moments of the 
$D\bar K^\ast$, $D^\ast \bar K$, and $D^\ast \bar K^\ast$ molecular states 
can be written as
\begin{align}
 \mathcal{R}_1(\mathrm{M^2},\mathrm{s_0})&= -(e_s-e_q)\frac{m_c^2}{2048 \pi^6} \Big[7 m_c^{10} I[-2] + 27 m_c^6 I[0] - 2 m_c^4 I[1] + 32 I[3]\Big]
    \nonumber\\
    &+ \frac{ \langle g_s^2 G^2\rangle \langle \bar q q \rangle }{  73728m_c \pi^4}  (e_s-e_q) \Big[ -12 m_{c}^2 \mathbb A[u_ 0] I[
   0] + (13 I_ 4[\mathcal S] + 
    12 (-I_ 4[\mathcal {\tilde S}] + 2
       I_ 6[\varphi_ {\gamma}])) (2m_{c}^2 I[0] - I[1]) 
       \Big]
    \nonumber\\
    & +\frac{ 13  \langle g_s^2 G^2\rangle f_{3 \gamma} }{147456 m_{c}^2 \pi^4} (e_s-e_q ) \Big[ (2 m_c^6 I[-1] + m_c^4 I[0] + I[2])  I_ 1[\mathcal V] \Big]
      \nonumber\\
    &+\frac{f_{3\gamma}}{2048 \pi^4} (e_s-e_q )  \Big[  (m_c^8 I[-1] + m_c^6 I[0] + 2 I[3]) I_ 1[\mathcal V]\Big], 
\end{align}
\begin{align}
 \mathcal{R}_3
 (\mathrm{M^2},\mathrm{s_0})&= \frac{m_{c}^2}{8192 \pi^6} 
 \Big[ 4 e_q (7 m_c^{10} I[-2] + 27 m_c^6 I[0] - 2 m_c^4 I[1] + 32 I[3]) - 
 e_c (m_c^{10} I[-2] + 32 m_c^8 I[-1] - 6 m_c^6 I[0] \nonumber\\
 &+ 
    40 m_c^4 I[1] - 3 m_c^2 I[2] + 64 I[3]) \Big]
    \nonumber\\
    &+ \frac{e_q\, \langle g_s^2 G^2\rangle \langle \bar q q \rangle }{  36864 m_{c} \pi^4} \Big[ 2 m_c^2 (6 \mathbb A[u_ 0] + 13 I_ 4[\mathcal S] + 
    12 I_ 4[\tilde {\mathcal S}]) I[
   0] - (13 I_ 4[\mathcal S] + 
    12 (I_ 4[\tilde {\mathcal S}] + 2 I_ 6[h_\gamma])) I[1] \Big]
    \nonumber\\
        & -\frac{ e_q\,  \langle g_s^2 G^2\rangle f_{3 \gamma} }{147456  m_{c}^2 \pi^4}\Big[  
 13  (2 m_c^6 I[-1] + m_c^4 I[0] + I[2]) I_ 1[\mathcal V] + 
 24 I_ 5[\psi^a] (2 m_c^8 I[-2] + m_c^4 I[0] - 4 m_c^2 I[1] + I[2]) \nonumber\\
 &+ 
 48 (2 m_c^8 I[-2] + 3 m_c^4 I[0] - 2 m_c^2 I[1] + I[2]) \psi^a[u_ 0]
    \Big]
    \nonumber\\
        &- \frac{e_q \, \langle g_s^2 G^2\rangle \langle \bar q q \rangle \chi}{  1536 m_{c} \pi^4}  \Big[   (2 m_c^6 I[-1] + m_c^4 I[0] + I[2]) \varphi_ {\gamma}[u_ 0]  \Big]
    \nonumber\\
        &+\frac{3 e_q\,m_c^3\, \langle \bar q q \rangle }{512 \pi^4}  \Big[ -2 m_c^6 I_ 6[h_\gamma] I[-2] - 
 m_c^4 \big (2 \mathbb A[u_ 0] + I_ 4[\mathcal  S] + 
    I_ 4[\tilde {\mathcal S}] + 4 I_ 6[h_\gamma] \big) I[-1] + 
 2 m_c^2 \big (\mathbb A[u_ 0] + I_ 4[\mathcal  S] \nonumber\\
  &+ 
    I_ 4[\tilde {\mathcal S}] + I_ 6[h_\gamma]\big) I[
   0] - (I_ 4[\mathcal  S] + I_ 4[\tilde {\mathcal S}]) I[1] \Big]\nonumber \\
          &-\frac{e_q\,f_{3\gamma}}{1024 \pi^4} \Big[  2  (m_c^{10} I[-2] + 2 m_c^8 I[-1] + m_c^6 I[0] + 
    4 I[3]) I_ 5[ \psi^a] + (m_c^8 I[-1] + 2 m_c^6 I[0] + 
    m_c^4 I[1] + 4 I[3]) I_ 1[\mathcal V] \nonumber\\
    &+ 
 4 m_c^6 (m_c^4 I[-2] - I[0])  \psi^a[u_ 0] \Big],  \\
    \mathcal{R}_5(\mathrm{M^2},\mathrm{s_0})&=  - \frac{m_{c}}{8192 \pi^6} 
 \Big[ 12 e_s \Big(7 m_c^{11} I[-2] + 27 m_c^7 I[0] - 2 m_c^5 I[1] + 
    32 m_c I[3] \Big) - 
 16 e_q \Big(7 m_c^{11} I[-2] + 21 m_c^{10} m_s I[-2] \nonumber \\
 &+ 27 m_c^7 I[0] + 
    45 m_c^6 m_s I[0] - 2 m_c^5 I[1] - 6 m_c^4 m_s I[1] + 
    32 m_c I[3] + 60 m_s I[3] \Big) + 
 e_c \Big(3 m_c^{11} I[-2] \nonumber\\
 &+ 96 m_c^9 I[-1] + 32 m_c^8 m_s I[-1] - 
    18 m_c^7 I[0] + 96 m_c^6 m_s I[0] + 120 m_c^5 I[1] + 
    96 m_c^4 m_s I[1] - 9 m_c^3 I[2] \nonumber\\
    &+ 32 m_c^2 m_s I[2] + 
    192 m_c I[3] + 256 m_s I[3]\Big) \Big]
    \nonumber\\
        &+ \frac{\langle g_s^2 G^2\rangle \langle \bar q q \rangle }{ 36864 m_{c} \pi^4} \Big[ 3 e_s m_c^2 (8 I_ 2[\mathcal S] + 11 I_ 2[\tilde {\mathcal S}]) I[0] +
  e_q (17 I_ 4[\tilde {\mathcal S}] - 
    15 I_ 4[ {\mathcal S}]) (-2 m_c^2 I[0] + I[1]) \Big]
    \nonumber\\
        & -\frac{ 11 \langle g_s^2 G^2\rangle f_{3 \gamma} }{147456  m_{c}^2 \pi^4}  (e_s - 2 e_q)\Big[  
  (2 m_c^6 I[-1] + m_c^4 I[0] + I[2])I_1[\mathcal V]
    \Big]
    \nonumber\\
    &-\frac{ e_q\,m_c^3\, \langle \bar q q \rangle }{512 \pi^4}  \Big[ 3 m_c^3 ((5 m_c - 4 m_s) I_ 4[\mathcal S] + 
    m_c I_ 4[\tilde {\mathcal S}]) I[-1] - 
 6 m_c ((3 m_c - 2 m_s) I_ 4[\mathcal S] + 
    m_c I_ 4[\tilde {\mathcal S}]) I[0] \nonumber\\
    &+ 
 3 m_c (m_c - m_s) I_ 1[\mathcal S] (m_c^4 I[-2] - 2 m_c^2 I[-1] + 
    I[0]) + (m_c^5 (5 m_c + 3 m_s) I[-2] - 
    6 m_c^3 (2 m_c + m_s) I[-1] \nonumber\\
    &+ 3 m_c (3 m_c + m_s) I[0] - 
    2 I[1]) I_ 1[\tilde {\mathcal S}] + 
 3 (I_ 4[\mathcal S] + I_ 4[\tilde {\mathcal S}]) I[1] \Big]\nonumber \\
&+\frac{ e_s\,m_c^2\, \langle \bar ss \rangle }{256 \pi^4}  \Big[ 3 m_c^7 (I_ 1[\mathcal S] + I_ 1[\tilde {\mathcal S}] + 
    2 (I_ 2[\mathcal S] + I_ 2[\tilde {\mathcal S}])) I[-2] - 
 6 m_c^5 (I_ 1[\mathcal S] + I_ 1[\tilde {\mathcal S} ] + 
    2 I_ 2[\tilde {\mathcal S}]) I[-1]  \nonumber\\
    &+ (3 m_c^3 (I_ 1[\mathcal S] + 
       I_ 1[\tilde {\mathcal S}] - 2 I_ 2[\mathcal S] + 
       2 I_ 2[\tilde {\mathcal S}]) + 
    32 (I_ 2[\mathcal S] + 3 I_ 2[\tilde {\mathcal S}])) I[0] \Big]\nonumber\\
          &+\frac{f_{3\gamma}}{3072 \pi^4} \Big[ (e_q m_c^3 (4 m_c^7 I[-2] - 3 m_c^5 I[-1] + 18 m_c^4 m_s I[-1] - 
      36 m_c^2 m_s I[0] + 7 m_c I[1] + 18 m_s I[1]) + 8 e_q I[3] \nonumber\\
      &+ 
   6 e_s (-m_c^{10} I[-2] + m_c^8 I[-1] + 2 m_c^6 I[0] + 
      2 I[3])) I_ 1[\mathcal V] \Big].
 \end{align}

\noindent Here, $\langle g_s^2 G^2 \rangle$ denotes the gluon condensate, while 
$\langle \bar{q} q \rangle$ and $\langle \bar{s} s \rangle$ represent the light ($u,d$) and strange quark condensates, respectively. 
The convolution integrals $I[n]$ and $I_i[\mathcal{A}]$ ($i=1,\ldots,6$), which encode the spectral and distribution-amplitude contributions entering the sum rules, are defined as follows:
\begin{align}
I[n]&= \int_{\mathcal M}^{s_0} ds \,s^n e^{-s/\mathrm{M^2}}\nonumber\\
 I_1[\mathcal{A}]&=\int D_{\alpha_i} \int_0^1 dv~ \mathcal{A}(\alpha_{\bar q},\alpha_q,\alpha_g)
 \delta'(\alpha_ q +\bar v \alpha_g-u_0),\nonumber
  \end{align}
  \begin{align}
   I_2[\mathcal{A}]&=\int D_{\alpha_i} \int_0^1 dv~ \mathcal{A}(\alpha_{\bar q},\alpha_q,\alpha_g)
 \delta'(\alpha_{\bar q}+ v \alpha_g-u_0),\nonumber\\
   I_3[\mathcal{A}]&=\int_0^1 du~ A(u)\delta'(u-u_0),\nonumber\\
  I_4[\mathcal{A}]&=\int D_{\alpha_i} \int_0^1 dv~ \mathcal{A}(\alpha_{\bar q},\alpha_q,\alpha_g)
 \delta(\alpha_ q +\bar v \alpha_g-u_0),\nonumber\\
   I_5[\mathcal{A}]&=\int D_{\alpha_i} \int_0^1 dv~ \mathcal{A}(\alpha_{\bar q},\alpha_q,\alpha_g)
 \delta(\alpha_{\bar q}+ v \alpha_g-u_0),\nonumber\\
 I_6[\mathcal{A}]&=\int_0^1 du~ A(u).\nonumber
 \end{align}
Here, $\mathcal M= (m_c+m_s)^2$,  $\mathcal{A}$ represents the photon DAs, and the measure ${\cal D}\alpha_i$ is given by
\begin{eqnarray}
\label{nolabel05}
\int {\cal D} \alpha_i = \int_0^1 d \alpha_{\bar q} \int_0^1 d
\alpha_q \int_0^1 d \alpha_g \,  \delta(1-\alpha_{\bar
q}-\alpha_q-\alpha_g)~.
\end{eqnarray}

\bibliography{DstarKbarstarmolecule.bib}
\bibliographystyle{elsarticle-num}

\end{document}